\documentclass[%
 reprint,
superscriptaddress,
nobibnotes,
 amsmath,amssymb,
aps,
]{revtex4-2}

\usepackage{graphicx}% Include figure files
\usepackage{dcolumn}% Align table columns on decimal point
\usepackage{bm}% bold math
\usepackage{lipsum}
\usepackage[capitalize]{cleveref}
\usepackage[dvipsnames]{xcolor}
\usepackage{comment}

\definecolor{internationalorange}{rgb}{1.0, 0.31, 0.0}

\definecolor{crimson}{rgb}{0.86, 0.08, 0.24}
\newcommand{\rev}[1]{#1}

\begin{document}

\title{Higher-order modeling of face-to-face interactions}% Force line breaks with \\

\author{Luca Gallo}
    \thanks{These authors contributed equally}
    \affiliation{
        ANETI Lab, Corvinus Institute for Advanced Studies (CIAS), Corvinus University, 1093, Budapest, Hungary
    }
    \affiliation{
        Department of Network and Data Science, Central European University, 1100 Vienna, Austria
    }
    \affiliation{
        Center for Social Data Science (SODAS), University of Copenhagen, 1353, Copenhagen, Denmark
    }

\author{Chiara Zappal\`a}%
    \thanks{These authors contributed equally}
    \affiliation{
        Center for Collective Learning, Corvinus Institute for Advanced Studies (CIAS), Corvinus University, 1093 Budapest, Hungary
    }
    \affiliation{
        Center for Social Data Science (SODAS), University of Copenhagen, 1353, Copenhagen, Denmark
    }

\author{Fariba Karimi}%
    \affiliation{
        Graz University of Technology, 8010 Graz, Austria
    }
    \affiliation{
        Complexity Science Hub, A-1080 Vienna, Austria
    }

\author{Federico Battiston}%
    \email{battistonf@ceu.edu}
    \affiliation{
        Department of Network and Data Science, Central European University, 1100 Vienna, Austria
    }

\date{\today}

\begin{abstract}
The most fundamental social interactions among humans occur face-to-face.
Their features have been extensively studied in recent years, owing to the availability of high-resolution data on individuals' proximity.
Mathematical models based on mobile agents have been crucial to understanding the spatio-temporal organization of face-to-face interactions.
However, these models focus on dyadic relationships only, failing to characterize interactions in larger groups of individuals.
Here, we propose a model in which agents interact with each other by forming groups of different sizes.
Each group has a degree of social attractiveness, based on which neighboring agents decide whether to join.
Our framework reproduces different properties of groups in face-to-face interactions, including their distribution, the correlation in their number, and their persistence in time, which dyadic models cannot replicate.
Furthermore, it captures homophilic patterns at the level of higher-order interactions, going beyond standard pairwise approaches.
Our work
\rev{provides further evidence that higher-order interactions are key to describe human face-to-face contacts}
%
%sheds light on the higher-order mechanisms at the heart of human face-to-face interactions, 
%
paving the way for further investigation of how group dynamics at a microscopic scale affects social phenomena at a macroscopic scale.
\end{abstract}

\maketitle

\section*{Introduction}
Despite the disruption caused by the COVID-19 pandemic and the rapid advances in telecommunication, face-to-face interactions still substantially influence our social life.
Research has recently called attention to the importance of face-to-face interactions across multiple frameworks, including creative idea generation and diffusion \cite{brucks2022virtual,duede2024being,lin2023remote}, information transfer \cite{yang2022effects,cahill2022covid}, education \cite{betthauser2023systematic,ochs2024learning,lichand2022impacts}, psychological well-being \cite{marinucci2022online}, and, naturally, epidemic disease spreading \cite{zhang2020changes}.
The availability of high-resolution data on face-to-face contacts has allowed researchers to uncover how those interactions display coherent spatio-temporal characteristics across several social contexts.
Specifically, face-to-face interactions show universal features, such as the absence of a characteristic scale for contact duration, the switching between low activity periods and high activity bursts, and considerable heterogeneity in interaction behaviors among individuals \cite{cattuto2010dynamics,stehle2011high,isella2011s,takaguchi2011predictability,fournet2014contact,stopczynski2014measuring,mastrandrea2015contact}.
The ubiquity of these features has thus posed the crucial challenge of explaining what mechanisms underlie their emergence.

Modeling frameworks based on mobile agents proved to be a valuable tool to study the organization of human face-to-face interactions.
In this scenario, agents move erratically in a spatial environment, and interactions occur every time they get close together \cite{sekara2016fundamental}. 
In addition, contacts between agents can be modulated by more complex mechanisms, including their attractiveness \cite{starnini2013modeling,starnini2016model}, their activeness and reachability \cite{zhang2016modelling}, their pairwise similarity \cite{starnini2016emergence,flores2018similarity}, or belonging to the same social group \cite{oliveira2022group}.
This class of models gave useful insights to understand the bursty \cite{starnini2013modeling} and small-world behavior of face-to-face interactions \cite{tang2010small}, as well as many social phenomena emerging from them, e.g. disease spreading \cite{frasca2006dynamical,buscarino2008disease}, spatial segregation and echo chamber formation \cite{starnini2016emergence}, and structural inequalities \cite{oliveira2022group}.

Those models, however, are limited as they describe face-to-face interactions only in terms of dyadic relationships between agents.
In fact, they adopt a temporal network representation \cite{holme2012temporal}, focusing on either the dynamics of dyadic contacts, i.e., links, \cite{starnini2013modeling}, or the mesoscopic level of social gatherings, i.e., connected components \cite{sekara2016fundamental,flores2018similarity}.
However, \rev{the fundamental structures of face-to-face contacts are many-to-many interactions rather than one-to-one interactions \cite{lehmann2023fundamental}.
Recent analyses of high-resolution face-to-face contacts shows indeed that} humans do not only interact in pairs but regularly engage in groups involving more than two individuals at the same time \cite{cencetti2021temporal}, a scenario that is better described by higher-order networks \cite{battiston2020networks}.
Recent works have investigated the higher-order nature of face-to-face interactions with models based on different social mechanisms, including self-reinforcement, memory, and attribute-related network mechanisms \cite{hoffman2020model,gallo2023higher,iacopini2023temporal}.
Yet, these higher-order network approaches do not incorporate the spatial mobility of agents, and current models based on mobile agents either overlook or fail to capture \cite{starnini2016model} the spatio-temporal features and the dynamical evolution of groups.

Here, we bridge this gap by introducing a model in which mobile agents interact with each other by forming groups of different sizes.
Each group is characterized by an intrinsic degree of social appeal that we call ``group attractiveness''.
Agents passing in the vicinity of a group choose whether to join it based on its attractiveness, while group members decide whether to stay or walk away.
We show how the Group Attractiveness Model (GAM) can reproduce different properties of groups in face-to-face interactions, including their statistics, the correlation in their number, and their temporal duration.
Furthermore, differently from low-order approaches, we demonstrate the potential of our model to correctly capture higher-order homophilic patterns not only at the level of pairwise contacts but also at that of groups.
Given its predictive power, the Group Attractiveness Model can foster the study of human face-to-face interactions, paving the way for further investigation of how group dynamics at a microscopic scale affects social phenomena at a macroscopic scale.

\section*{Results}
\subsection*{The Group Attractiveness Model}
\begin{figure}[t!]
    \centering
    \includegraphics[width=\columnwidth]{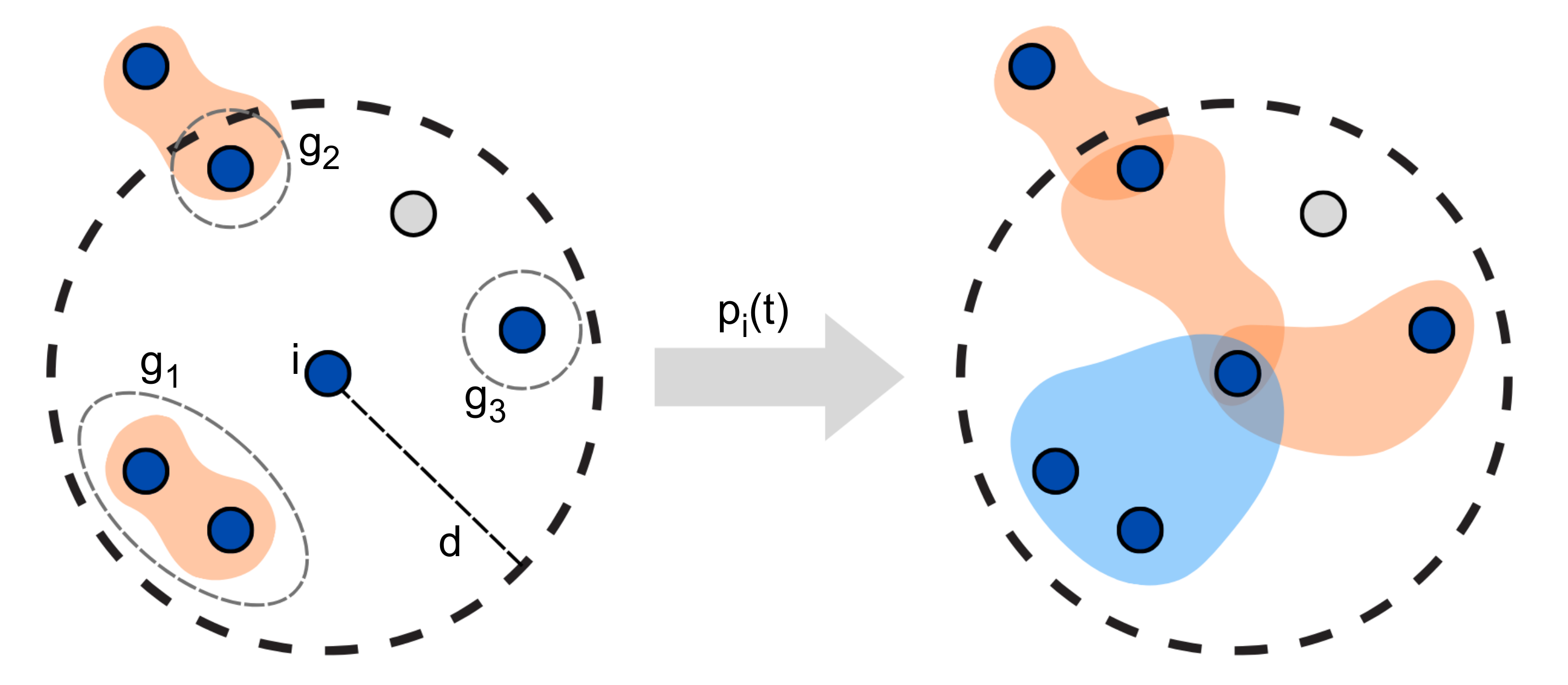}
    \caption{\textbf{Schematic illustration of the Group Attractiveness Model}. At each time step $t$, each active agent $i$ (blue) considers the groups lying within a radius $d$ from it, and interacts with all of them with a probability $p_i(t)$ that depends on the mean attractiveness of the neighboring groups. The agent interacts only with the group members within its scope and ignores inactive nodes (gray). With the complementary probability $1 - p_i(t)$, the agent moves away in a random direction, with a step of length $v$.}
    \label{fig:1}
\end{figure}

In the Group Attractiveness Model (\cref{fig:1}), $N$ agents are placed in a square environment of size $L\times L$, with periodic boundary conditions\rev{, i.e., when an agent crosses the left (top) boundary of the square, it reappears at the right (bottom) boundary}. 
Each agent $i$ has a value of attractiveness, $a_i$, which represents how appealing the agent is to the others.
\rev{Following \cite{starnini2013modeling,starnini2016model}, we operationalize attractiveness as the power of individuals to make others willing to interact with them, without making any assumptions on the social, economic, or behavioral factors that may determine it.}
The value of attractiveness is sampled from a uniform distribution in the interval $[0,1]$.
Agents can be isolated, or they can be part of a group.
An agent can decide to interact with the groups surrounding it or to walk away in a random direction.
How attractive a group is derives from how attractive its members are.
Formally, for each group of agents $g$, we define its attractiveness as
\begin{equation}
    a_g = \prod\limits_{j\in g} a_j,
    \label{eq:GA}
\end{equation}
where index $j$ runs over all agents forming group $g$.
Note that an isolated agent constitutes a group of one member (group $g_3$ in \cref{fig:1}).
Since the attractiveness of an individual does not depend on that of others, the average attractiveness of a group of size $s$ is $\overline{a}^s$, with $\overline{a}$ denoting the average individual attractiveness.
%We remark that a group's attractiveness is smaller than its members' individual attractiveness, i.e., $a_g < a_j, \, \forall j\in g$.
Consequently, the group attractiveness scales nonlinearly with its size, with larger groups being, on average, less attractive than smaller ones.
%Such modeling choice finds motivation in previous results on human face-to-face interactions, which highlight how large groups are more unstable than small ones \cite{stehle2010dynamical,gallo2023higher}, due to the higher propensity of individuals to leave them \cite{iacopini2023temporal}, a phenomenon known as schisming \cite{egbert1997schisming}.
At time step $t$, each agent $i$ considers the \rev{the set of groups, $\mathcal{N}(i)$,} located within a distance $d$ from it and interacts with all of them with probability
\begin{equation}
    p_i(t) = \frac{1}{|\mathcal{N}(i)|}\sum\limits_{g\in\mathcal{N}(i)} a_g,
    \label{eq:eq2}
\end{equation}
%where $\mathcal{N}(i)$ indicates the set of groups in the vicinity of $i$.
Therefore, when an agent chooses to interact with a group of size $s$, a group of size $s+1$ is formed at time $t+1$.
Groups in our model thus change gradually, with the addition of one member at a time, a mechanism supported by evidence from real-world human face-to-face interactions \cite{cencetti2021temporal,gallo2023higher,iacopini2023temporal}.
%We remark that, 
If $i$ is already part of a group, we consider the group formed by the other members to be part of $\mathcal{N}(i)$.
Therefore, when $i$ decides to interact, it rejoins the group it was part of, i.e., the group persists in time. 
Also, when a group partially lies within the scope of an agent, the latter interacts only with those agents that are at a distance smaller than $d$ (e.g., only one member of group $g_2$ in \cref{fig:1} is close to agent $i$, so a group of size 2 is formed between them).
%Also, when an agent is already part of a group, it simply keeps interacting with its members. 
When an agent does not interact with its neighboring groups, it makes a step of length $v$ along a direction given by a randomly chosen angle $\xi\in [0,2\pi]$, leaving all the groups it was part of in the previous time step.
Hence, a group persists in time only when all its members decide to interact with the others \cite{stehle2010dynamical}.
Finally, we assume that agents can be active or inactive.
While an active agent can walk or form groups with other agents, an inactive agent neither moves nor interacts with the others.
Inactive agents can become active with probability $r_i$, which we sample from a uniform distribution in $[0,1]$, while active agents that are isolated can become inactive with probability $1 - r_i$.
\rev{This further mechanism models the empirical observation that people can leave or join the social context \cite{starnini2013modeling} and allows us to distinguish between the individuals that are not in the system from those that are in the system but do not engage in group conversations.}

\rev{At the end of each time step, the system consists of interacting groups that potentially overlap, i.e., share one or more members, plus a set of agents that do not interact.
From a network science perspective, this can be described as a hypergraph \cite{battiston2020networks}, which evolves over time \cite{cencetti2021temporal}.}

\rev{The attractiveness of an agent represents the likelihood that other agents will interact with it when passing by.
This likelihood can be influenced by multiple factors, e.g., the fame of an individual, and it may be hard to measure empirically.
To avoid introducing any biases, we sample the attractiveness from a uniform distribution.
Still, the dynamics of the system may be robust under other choices, such as a normal distribution (see Figs.~S1-S2 in the Supplementary Information). 
}

In the Group Attractiveness Model, three different social processes, i.e., joining pre-existing group conversations, remaining in them, and leaving them, are regulated by the average attractiveness of the groups surrounding them.
The attractiveness of those groups decrease with their size.
Previous research motivates this modeling choice:
On the one hand, studies have shown that larger groups involved in exclusive interactions, e.g., conversations, tend to be more impermeable and more repelling, hence less attractive, to passersby \cite{knowles1973boundaries,knowles1976group,zander1979psychology,mullen1991boundaries}.
On the other hand, there is an upper limit on how many individuals can converse \cite{dunbar1995size}, after which the conversation splits up into two or more conversations \cite{egbert1997schisming}, a phenomenon known as schisming \cite{sacks1974simplest}.

%Naturally, the attractiveness of conversational groups depends on the social context and nature of groups, which our model does consider.
%Also, group attractiveness could influence the decision of individuals to join, leave, or remain in a group in different ways.
%Despite these simplifying assumptions, the Group Attractiveness Model successfully replicates key features of group face-to-face interactions, as we will show in the next sections.

\begin{figure*}[t!]
    \centering
    \includegraphics[width=\textwidth]{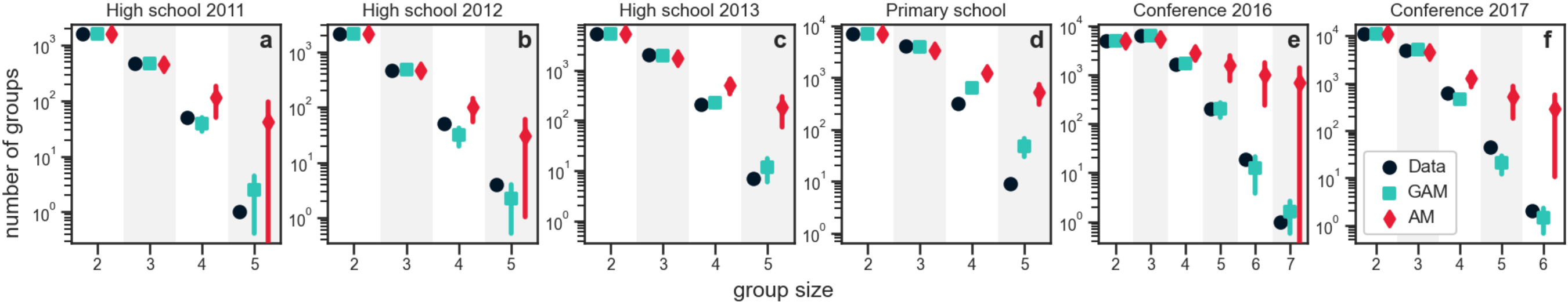}
    \caption{\textbf{The GAM reproduces the empirical group statistics}. 
    Panels \textbf{a} to \textbf{f} report the distribution of unique groups of different sizes in a given social system (black circles), as well as the predictions of the Group Attractiveness Model (blue squares) and the Attractiveness Model \cite{starnini2013modeling} (red diamonds). 
    Markers represent the average number of groups generated by the models over 100 simulations, while error bars indicate the standard deviation. 
    Whereas the Attractiveness Model largely overestimates the number of large groups in the data, the Group Attractiveness Model correctly predicts the group statistics.}
    \label{fig:2}
\end{figure*}

\subsection*{Higher-order statistics of human face-to-face interactions}
To test the capability of the Group Attractiveness Model to reproduce higher-order patterns of human face-to-face interactions, we analyze \rev{thirteen high-resolution datasets, ten coming from the SocioPatterns collaboration \cite{cattuto2010dynamics}, two from as many experiments in Utah \cite{toth2015role}, and one from the Copenhagen Network Study \cite{sapiezynski2019interaction}}.

\rev{SocioPatterns data report dyadic contacts between individuals facing each other within a 1–1.5 meter range with a high temporal resolution, i.e., 20-second intervals, recorded using Radio Frequency Identification (RFID) devices.
The datasets describe the dynamics of contacts between individuals in different social contexts, specifically a primary school (``PS'') \cite{stehle2011high}, a high school (``HS11'', ``HS12'', and ``HS13'') \cite{fournet2014contact,mastrandrea2015contact}, two scientific conferences (``C16'' and ``C17'') \cite{genois2019building}, an art exhibition (``SG'') \cite{isella2011s}, a hospital (``H'') \cite{vanhems2013estimating}, a village in Malawi (``MV'') \cite{ozella2021using}, and a workplace (``WP'') \cite{genois2018can}.
Besides contact patterns, the datasets on schools and conferences (six in total) contain information about participants' gender.}

\rev{Datasets from the Utah experiments report contacts between pairs of people located 2 meter or less from each other, with a 20-second resolution, in a middle school (``MS'') and an elementary school (``ES''), recorded with Wireless Ranging-Enabled Nodes (WRENs).
Data from the Copenhagen Network Study consists of dyadic interactions in a university campus (``UC''), registered every 5 minutes using the Bluetooth technology, which has a range of 10 meters. 
Here we focus on the six SocioPatterns datasets with gender, and report the analysis of the other six systems in the Supplementary Information.
}

As all datasets store interactions as dyadic contacts, we first reconstruct group interactions, i.e., face-to-face conversations involving two or more individuals, leveraging the fine-grained temporal information of the data.
Specifically, if at a time $t$ we find all possible dyads among $s$ individuals, we assume that they are part single group of size $s$ (see Materials and Methods for details).
The statistics of unique groups of different sizes for the ten datasets are shown in \cref{fig:2} and Fig.~S1 as black circles.
\rev{In general, smaller groups are more abundant than larger ones, except for the ``C16'' and the ``UC'' dataset (see Table~S1 in Supplementary Information).}

We now want to verify whether our model is able to reproduce the distribution of unique groups in those social systems.
We initialize the model simulation by randomly placing each agent in the environment and setting agents active with probability $1/2$.
We fix $v=d=1$ and the number of agents $N$ as the number of individuals forming the largest connected component in the hypergraph of contacts (see Supplementary Information for more details).
The simulation stops once the number of groups of size two generated reaches the empirical value.
The number of groups of size three or more utterly depends on the agent density.
For instance, if the size $L$ of the environment is significantly large, then agents would rarely get in contact with each other, making the formation of large groups quite unlikely.
On the other hand, when agents are close to each other, i.e., when the density is high, it is more likely that agents form groups of various sizes.
Hence, we fit the value of $L$ that best reproduces the group statistics in the dataset (see Supplementary Information for the best-fit values of $L$ in each system).

As a comparison, we consider the (individual) Attractiveness Model (AM) proposed in \cite{starnini2013modeling}.
Since such a model accounts for groups of two agents only, we extract groups of larger size following the same procedure adopted for empirical data.
We then fit the size environment $L$ in the same way as for the GAM.
For both models, we run 100 simulations and consider the average number of groups of different sizes, as well as the standard deviation as an estimator of the model variability.

\begin{figure}[t!]
    \centering
    \includegraphics[scale=0.45]{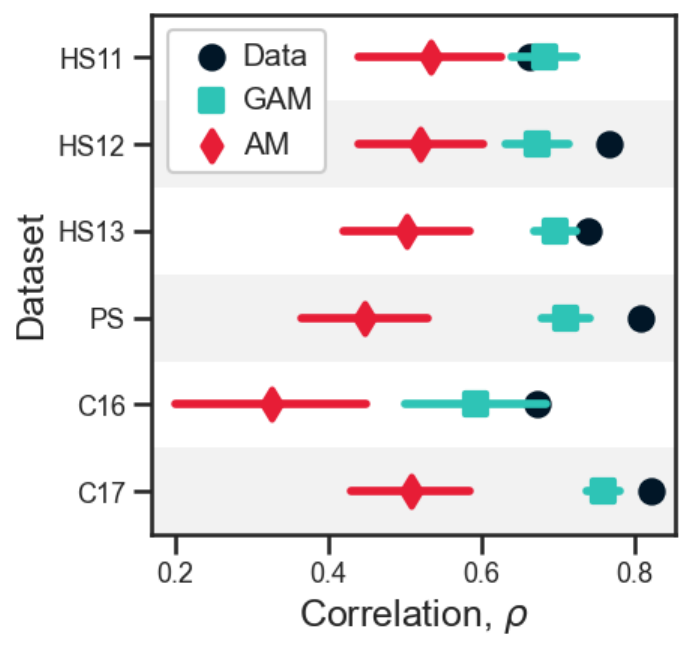}
    \caption{\textbf{The GAM reproduces the correlation between the number of groups of sizes two and three}. Correlation in the number of groups of size two and three in various social contexts (black circles) is compared to the predictions of the Group Attractiveness Model (blue squares) and the Attractiveness Model (red diamonds). 
    Markers represent the average correlation over 100 simulations, while error bars indicate the standard deviation. 
    The Attractiveness Model systematically underestimates correlations in the number of groups, while the Group Attractiveness Model better reproduces empirical values of correlation.}
    \label{fig:3}
\end{figure}

In \cref{fig:2} and Fig.~S1, we show the average number of unique groups predicted by the GAM (blue squares) and the AM (red diamonds), other than the group statistics of the datasets (black circles).
%In \cref{fig:2}, we show the group statistics of six datasets (black circles), as well as the average number of groups predicted by the GAM (blue squares) and the AM (red diamonds).
In general, the Group Attractiveness Model reproduces the distributions of unique groups of different sizes; instead, the (individual) Attractiveness Model significantly overestimates the number of larger groups.
In those cases where the GAM is not able to capture the exact group statistics (e.g., ``PS'' dataset), we still observe a better agreement compared to the AM (see \rev{Fig.~S3 and Fig.~S5} for a more detailed analysis of the additional datasets).
This result highlights the need to consider group attractiveness to properly model non-dyadic face-to-face interactions.

Next, we aim to understand whether individuals participating in groups of a given size also participate in groups of a different size.
The presence of those correlations, and in particular the empirical tendency of face-to-face group interactions to be nested (i.e., individuals interacting in a group at a given time also interact in subgroups at other times)\cite{lotito2022higher,landry2024simpliciality}, can have important consequences, for example promoting contagion dynamics \cite{landry2020effect,larock2023encapsulation,kim2023contagion}.
Motivated by this, we examine the capability of the Group Attractiveness Model to reproduce correlations in the number of groups, focusing on groups of sizes two and three.

We count the unique number of groups of size two, $k^{(2)}_i$, and three, $k^{(3)}_i$, in which each individual $i$ takes part at any moment within the observation time of the system, and evaluate the Pearson correlation coefficient, $\rho$, between these two quantities.
We then simulate 100 times the Group Attractiveness Model, using the parameters fitted from the group size distributions, and evaluate for each run the linear correlation between the groups of sizes two and three. 
Again, we consider the Attractiveness Model as a reference model.

The results are reported in \cref{fig:3}, \rev{Fig.~S4 and Fig.~S6}.
We observe that, in general, face-to-face interactions in pairs and triads tend to be highly correlated in real-world systems (black circles), with $\rho$ varying between 0.56 and 0.85.
This aspect of the empirical datasets is well reproduced by the GAM (blue squares), for which the average correlation coefficient never goes below 0.54.
Moreover, the GAM is able to predict the exact value of $\rho$ for half of the systems while slightly underestimating it for the others.
The AM, instead, systematically underestimates the correlations between groups of sizes two and three (red diamonds), with values consistently below 0.53, down to 0.33.

The correlation analysis \rev{highlights the relevance of considering higher-order mechanisms to describe} human face-to-face interactions. 
In the Attractiveness Model, which is based on a dyadic approach, groups of more than two agents are constructed as a collection of pairs of agents. 
For example, if three agents form three pairs at a given time step, we assume them to interact in a group of three agents.
Consequently, the correlation between groups of two and three agents reduces, especially in those scenarios with a high density of agents, i.e., the ``C16'' dataset.
The lower correlation in the AM is not simply an effect of how we reconstruct groups from pairwise interactions, as in empirical systems, where groups are obtained in the same way, we observe high correlation values.
This means that groups in real-world systems are not simply a collection of dyadic contacts.
In fact, the Group Attractiveness Model, which naturally accounts for group interactions, correctly reproduces the high values of correlation observed in the data.

\subsection*{The hierarchical structure of group burstiness}
A distinctive characteristic of human face-to-face interactions is their bursty behavior \cite{cattuto2010dynamics,karsai2012universal}.
In particular, the duration of contacts between individuals displays broad-tailed distributions, indicating that most contacts are brief and few last for long periods of time, with no characteristic time scale.
Recent literature has shown that burstiness is not limited to pairwise contacts, as group interactions show similar temporal patterns \cite{zhao2011social,cencetti2021temporal,iacopini2023temporal}.
Interestingly, the distributions of the contact duration are typically organized in a hierarchy, with smaller groups showing broader distributions compared to larger ones, a feature that also emerges in the social systems we investigate (here we focus on the ``HS11'' dataset, panel \textbf{a} of \cref{fig:4}; the analysis of the other datasets is reported in the Supplementary Information, Figs.~S3-S11).
Note that here we define the contact duration as the number of consecutive time steps for which an interaction is present. 

Whereas the (individual) Attractiveness Model succeeds in reproducing the broad-tailed distribution of pairwise contacts \cite{starnini2013modeling}, it fails to recover the hierarchical structure of group interactions.
In particular, the model predicts larger groups to be more stable than smaller ones, namely that groups with more individuals remain in contact for longer (panel \textbf{c} of \cref{fig:4} and Figs.~S3-S11 in Supplementary Information) \cite{starnini2016model}.
This discrepancy with empirical evidence is probably due to the fact that larger groups of agents in the AM tend to be more attractive, as it is more likely that individuals with high attractiveness are members of the group. 

Contrarily, capitalizing on the higher-order mechanism of group formation based on group attractiveness, the GAM is able to produce broad-tailed distributions for the contact duration as well as their hierarchical organization (panel \textbf{b} of \cref{fig:4} and \rev{Figs.~S7-S18} in the Supplementary Information).
Yet, we observe that the distributions are often narrower compared to the empirical ones, especially in scenarios where the density of agents is high (see results on the ``C16'' dataset in the Supplementary Information).
This is probably due to how we define the probability that an agent interacts with its neighbors, i.e., \cref{eq:eq2}.
Specifically, in a dense environment, each agent will interact with a probability that tends to the average value of the group attractiveness, meaning that individuals with high attractiveness, namely those contributing the most to the persistence of interactions, do not have a strong effect.
Further studies should aim to understand the relationship between the broadness of the distributions and their hierarchical organization.

\begin{figure}[t!]
    \centering
    \includegraphics[width=\columnwidth]{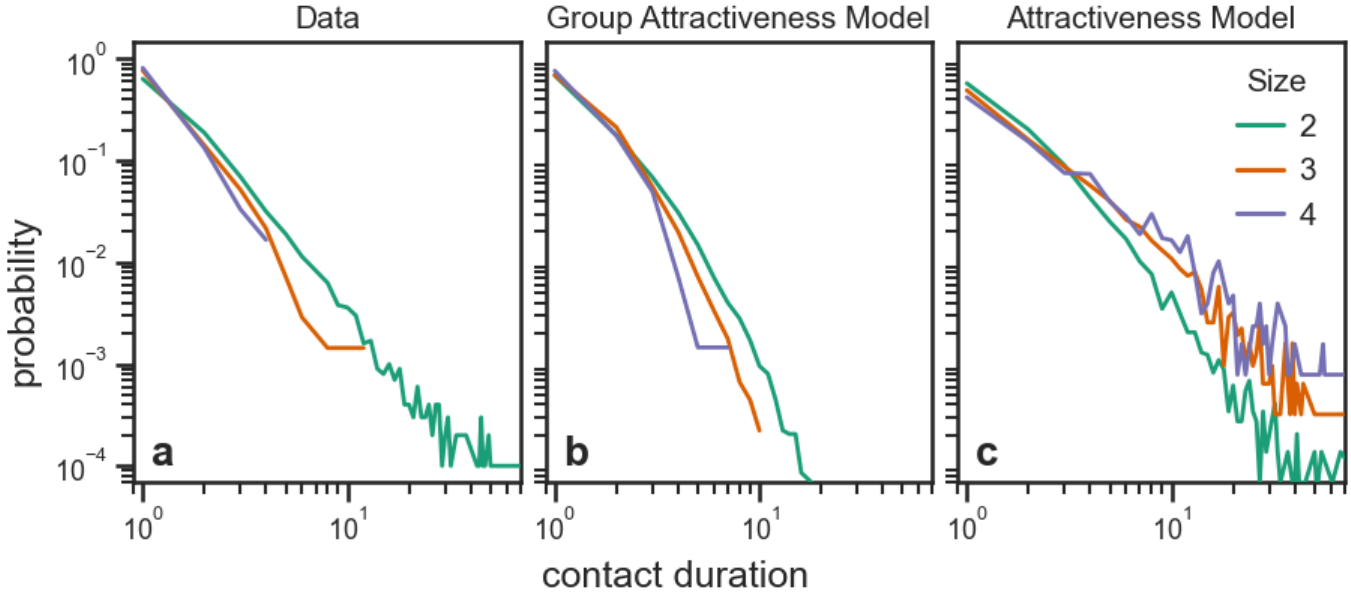}
    \caption{\textbf{Hierarchical organization of higher-order burstiness}. We show the distributions of contact duration for groups of different sizes in the ``HS11'' dataset (panel \textbf{a}), as well as those predicted by the Group Attractiveness Model (panel \textbf{b}) and the Attractiveness Model (panel \textbf{c}).
    The Attractiveness Model predicts larger groups to be more stable than smaller ones.
    Instead, the Group Attractiveness Model correctly reproduces the hierarchical organization of the distributions observed in the data, as groups with fewer individuals remain in contact longer than groups with more individuals.}
    \label{fig:4}
\end{figure}

\begin{figure*}[t!]
    \centering
    \includegraphics[width=.8\textwidth]{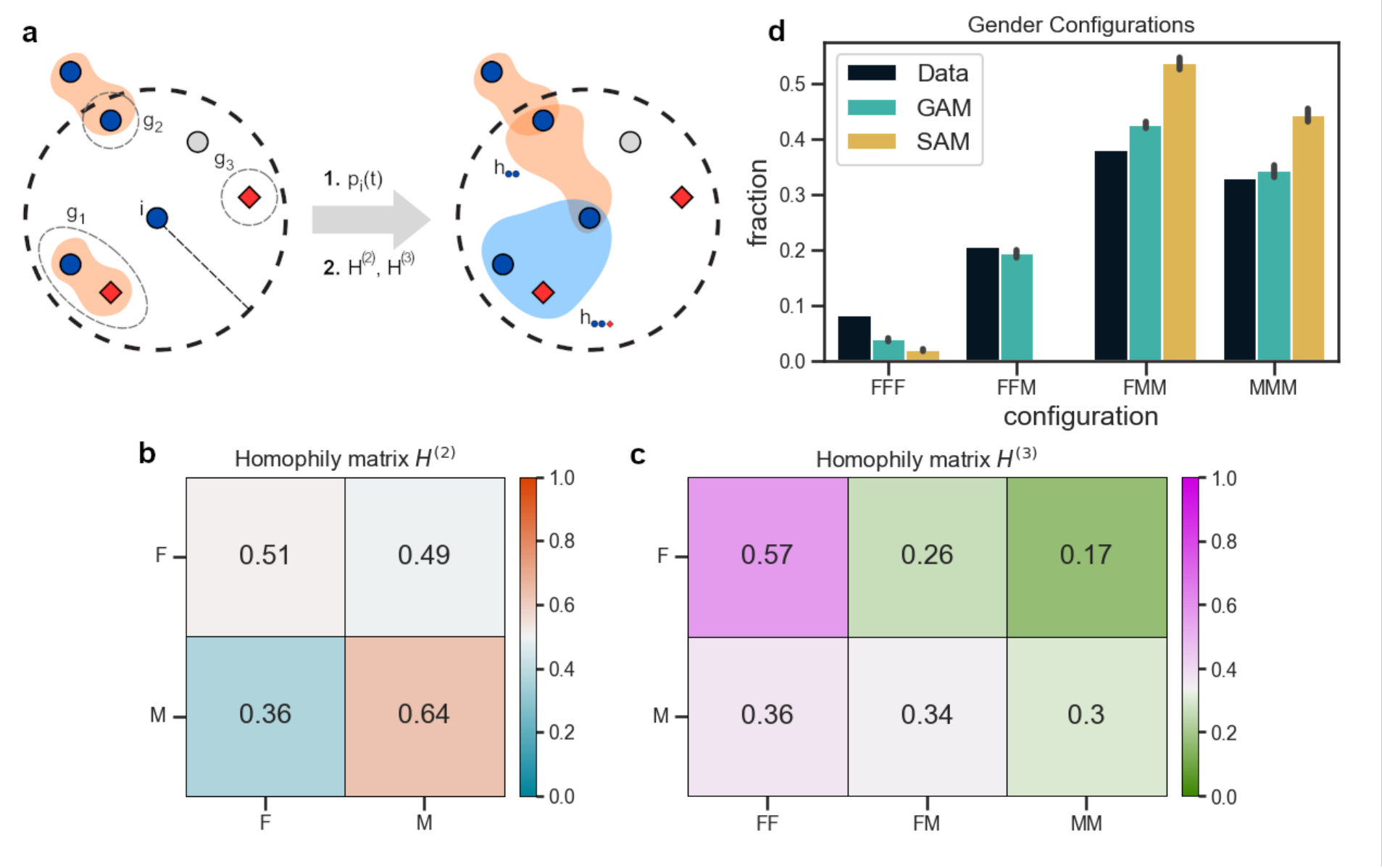}
    \caption{\textbf{Higher-order homophily in face-to-face interactions}. 
    Panel \textbf{a} shows a schematic of the Group Attractiveness Model with homophily. 
    At each time step $t$, an active agent $i$ (blue) considers the groups within its scope and decides to interact with a probability $p_i(t)$ based on the mean attractiveness of the neighboring groups (see \cref{eq:eq2}); 
    if it stays, the agent chooses the group(s) to which it connects based on the homophily matrices $H^{(2)}$, $H^{(3)}$, and so on, which depend on its own attributes (shapes and colors) and those of the group member(s).
    Panels \textbf{b} and \textbf{c} display the homophily matrices $H^{(2)}$ and $H^{(3)}$, modulating the formation of groups of two and three individuals, respectively, obtained for the interactions in the ``HS11'' dataset.
    \rev{The element $(i,j)$ of $H^{(2)}$ is the probability that an agent with attribute $i$ (either F or M) interacts with a neighbor with attribute $j$ (either F or M), while the element $(i,k)$ of $H^{(3)}$ is the probability that an agent with attribute $i$ interacts with a neighboring group of two agents whose attributes are in state $k$ (either FF, FM, or MM).}
    Men prefer to interact with other men at the level of pairwise interactions \rev{($h^{(2)}_{MM}>h^{(2)}_{MF}$)}, while women are more homophilic in groups of three individuals \rev{($h^{(3)}_{FFF}>h^{(3)}_{FFM}$ and $h^{(3)}_{FFF}>h^{(3)}_{FMM}$)}.
    Panel \textbf{d} shows the fraction of unique groups of size three in the different gender configurations present in the ``HS11'' dataset (black bars), together with those predicted by the Group Attractiveness Model (blue bars) and by the Social-Attractiveness Model \cite{oliveira2022group} (yellow bars).
    The value of the bars represents the average fraction over 100 simulations, while the error bars indicate the standard deviation. 
    While the Social-Attractiveness Model overestimates the tendency of men to interact with their same gender, the Group Attractiveness Model is in good agreement with the empirical mixing patterns.
    }
    \label{fig:5}
\end{figure*}

\subsection*{Higher-order homophily in face-to-face interactions}
In many social contexts, people prefer to build ties with others whom they perceive as similar to themselves \cite{mcpherson2001birds}. 
This pervasive characteristic, known as homophily, shapes the ``social world'' of individuals, thus profoundly influencing how behavior spreads \cite{christakis2007spread}, biases \cite{lee2019homophily} and social norms \cite{centola2005emperor} form, and segregation emerges \cite{schelling1971dynamic,currarini2009economic}.
Homophily characterizes face-to-face interactions as well \cite{stehle2011high,stehle2013gender,ozella2021using}, driving the onset of inequalities even at such a fundamental scale \cite{oliveira2022group}.

While homophily is usually measured at the level of pairs of individuals, recent studies have aimed to capture it at the level of groups of three or more individuals \cite{veldt2023combinatorial,sarker2024higher,rizi2025homophily}.
We can use the Group Attractiveness Model to analyze higher-order homophilic patterns in face-to-face interactions.
Specifically, we enrich the model by associating agents with a set of attributes and by tuning the probability that an agent interacts with its neighbors according to their attributes.
\rev{Following \cite{oliveira2022group}, we define} group formation as a two-step process that incorporates attractiveness and homophilic preferences: 
First, each agent decides whether to stay or to walk away based on the attractiveness of its neighborhood (see \cref{eq:eq2});
If it stays, the agent chooses the group(s) to which it connects based on its own attributes and those of the group member(s) (panel \textbf{a} of \cref{fig:5}).

To illustrate the second step, let us assume that each agent is associated with a single attribute.
An agent with attribute $\alpha$ close to an agent with attribute $\beta$ will form a group of two with probability $h^{(2)}_{\alpha\beta}$.
Note that $h^{(2)}_{\alpha\beta}$ represents the probability that it is the agent with attribute $\alpha$ to start the interaction, and in general $h^{(2)}_{\alpha\beta} \neq h^{(2)}_{\beta\alpha}$.
Similarly, if the agent is close to a group of two agents having attributes $\beta$ and $\gamma$, respectively, it will form a group of three with probability $h^{(3)}_{\alpha\beta\gamma}$. 
Therefore, the probability of forming groups of various sizes based on the agents' attributes is determined by a set of homophily matrices, $H^{(2)}$, $H^{(3)}$, and so on.
Here, we will focus on a single binary attribute, i.e., $\alpha\in\{0,1\}$, using the information on gender contained in six of the datasets to test the ability of our model to reproduce higher-order mixing patterns in face-to-face interactions (from now on, attribute $0$ will denote women, while attribute $1$ will denote men).

To determine the elements of the homophily matrix $H^{(2)} = [[h_{00},h_{01}],[h_{10},h_{11}]]$ (superscripts are dropped for simplicity), we evaluate the fraction of groups of size two in the different configurations, i.e., female-female, male-male, and female-male, that are formed from two individuals not previously interacting, that is $i$ and $j$ form a group at time $t$ but they are not part of any common group at time $t-1$.
Such fractions can be written in terms of the elements of the homophily matrix (see Materials and Methods for details), as 
\begin{eqnarray}
    e_{00} = \frac{f_0^2 (1-h_{01}^2)}{f_0^2 (1-h_{01}^2) + 2f_0f_1 (1-h_{00}h_{11}) + f_1^2 (1-h_{10}^2)}, \nonumber \\
    e_{01} = \frac{2f_0f_1 (1-h_{00}h_{11})}{f_0^2 (1-h_{01}^2) + 2f_0f_1 (1-h_{00}h_{11}) + f_1^2 (1-h_{10}^2)}, \\
    e_{11} = \frac{f_1^2 (1-h_{10}^2)}{f_0^2 (1-h_{01}^2) + 2f_0f_1 (1-h_{00}h_{11}) + f_1^2 (1-h_{10}^2)}, \nonumber
\end{eqnarray}
where $e_{00}$, $e_{01}$, and $e_{11}$, denotes the fractions of groups formed by two women, a woman and a man, and two men, respectively, while $f_0$ and $f_1 = 1-f_0$ represent the fraction of women and men.
To estimate the elements of the homophily matrix $H^{(3)} = [[h_{000},h_{001},h_{011}],[h_{100},h_{101},h_{111}]]$, we count the groups of size three in the different configurations that are formed by aggregation of an individual in a group of size two, i.e., at time $t-1$ two individuals $i$ and $j$ form a group, at time $t$ an individual $k$, not previously interacting with them, joins the group.
\rev{In other words, we count the transitions of dyads to triads driven by the addition of one member.}
Based on the gender of the individuals joining the group, we have two sets of transitions.
A woman can join a group of two other women, two men, or a woman and a man.
The fraction of these transitions can be written in terms of the first row of the homophily matrix $H^{(3)}$ (see Materials and Methods), namely
\begin{eqnarray}
    \tau_{0\rightarrow (0,0)} = \frac{\varepsilon_{00}h_{000}}{\varepsilon_{00}h_{000} + \varepsilon_{01}h_{001} + \varepsilon_{11}h_{011}}, \nonumber \\
    \tau_{0\rightarrow (0,1)} = \frac{\varepsilon_{01}h_{001}}{\varepsilon_{00}h_{000} + \varepsilon_{01}h_{001} + \varepsilon_{11}h_{011}}, \\
    \tau_{0\rightarrow (1,1)} = \frac{\varepsilon_{11}h_{011}}{\varepsilon_{00}h_{000} + \varepsilon_{01}h_{001} + \varepsilon_{11}h_{011}}, \nonumber
\end{eqnarray}
where $\tau_{0\rightarrow (\alpha,\beta)}$ indicates the fractions of transitions, while  $\varepsilon_{\alpha\beta}$ denotes the fractions of unique groups of size two in the various configurations. 
\rev{Note that $e_{\alpha\beta}$ indicates dyads emerging at a given time from two not previously interacting individuals; instead, $\varepsilon_{\alpha\beta}$ denotes dyads that are active at a given time, including those formed in previous time steps that persisted over time, and those generated from other dynamics, e.g., a group of three that loses a member.}
In the same way, men can join groups of two individuals in different configurations, the fraction of which can be expressed in terms of the second row of the homophily matrix $H^{(3)}$, namely
\begin{eqnarray}
    \tau_{1\rightarrow (0,0)} = \frac{\varepsilon_{00}h_{100}}{\varepsilon_{00}h_{100} + \varepsilon_{01}h_{101} + \varepsilon_{11}h_{111}}, \nonumber \\
    \tau_{1\rightarrow (0,1)} = \frac{\varepsilon_{01}h_{101}}{\varepsilon_{00}h_{100} + \varepsilon_{01}h_{101} + \varepsilon_{11}h_{111}}, \nonumber \\
    \tau_{1\rightarrow (1,1)} = \frac{\varepsilon_{11}h_{111}}{\varepsilon_{00}h_{100} + \varepsilon_{01}h_{101} + \varepsilon_{11}h_{111}}.
\end{eqnarray}
A similar approach can be adopted to evaluate the matrices modulating the formation of groups of four or more individuals \rev{(see Methods)}.
As larger groups are less abundant, for simplicity, we here limit our analysis to groups of size two and three.

Panels \textbf{b} and \textbf{c} of \cref{fig:5} display the homophily matrices $H^{(2)}$ and $H^{(3)}$ obtained for the interactions in the ``HS11'' dataset (see Supplementary Information for the analysis of the other five systems).
At the level of pairwise interactions, we observe that women do not have a clear homophilic behavior, as they interact with other women and men with almost the same probability.
Conversely, men are strongly homophilic, as the model predicts a substantial difference between the interaction probabilities.
Remarkably, things change in groups of size three.
In this case, women tend to be more homophilic, while men do not have a strong gender preference when joining groups of two individuals. 
Homophilic preferences depend on the group size in a nontrivial way: Here we observe discordant behavior, i.e., men tend to be homophilic in pairs, whereas women in triples, while other social systems can display a consistent pattern (see Panels \textbf{a} and \textbf{b} of \rev{Figs.~S19-S23} in Supplementary Information).

Finally, we test the capability of the GAM to reproduce mixing patterns in social systems.
Panel \textbf{d} of \cref{fig:5} shows the fraction of unique groups of size three in the different gender configurations present in the data (black bars), together with those predicted by the GAM (blue bars).
For comparison, we consider the Social-Attractiveness Model (SAM) proposed in \cite{oliveira2022group} (yellow bars). 
Similarly to our model, in the SAM a population of mobile agents performs a random walk interacting with the others based on the intrinsic attractiveness of individuals and their attributes, i.e., gender.
Yet, this model only accounts for pairwise interactions, so mixing patterns at the level of groups of three agents are ultimately determined by homophily at the level of pairs.
\rev{Note again that the fraction of unique triads formed over time differs from the fraction of triads generated at a given time step as a result of an agent joining a dyad, i.e., $\tau_{0\rightarrow (\alpha,\beta)}$.}
Our results show that considering higher-order homophily better reproduces the gender configurations in the data.
Particularly, we observe that the SAM overestimates the tendency of men to interact with their same gender, generating too many groups with three men or two men and a woman.
Conversely, our model provides significantly better predictions of the empirical mixing patterns (the better performance of GAM is consistent across different datasets; see Panels \textbf{c} of \rev{Figs.~S19-S23} in Supplementary Information).
Still, we observe some mismatch, particularly in the fraction of groups with three women.
This might have several explanations, including differences in individual behavior or in the frequency of contacts between the two genders \cite{stehle2013gender}, as well as other mechanisms of group formation/dissolution that the Group Attractiveness Model does not account for \cite{cencetti2021temporal,gallo2023higher,iacopini2023temporal}.
Overall, our results shed light on how higher-order effects can influence mixing patterns in social systems, underlining the importance of measuring homophily at the level of groups.

\section*{Discussion}
Humans are social animals that communicate, gather, and live in groups.
Even a fundamental level of social interactions, such as face-to-face contacts, is characterized by groups of various sizes.
Although models based on dyadic representations, i.e., complex networks, have proved to be a valuable tool to characterize different properties of face-to-face interactions, they fall short when it comes to replicating structural and temporal features of groups. 
This is crucial, as group interactions can dramatically change the collective behavior of complex social systems, leading to super-exponential disease spreading \cite{st2021universal}, triggering critical mass effects in social contagion \cite{iacopini2019simplicial,de2020social}, and boosting the ability of committed minorities \cite{centola2018experimental} to overturn social norms \cite{iacopini2022group}.

In this paper, we presented the Group Attractiveness Model, an agent-based model that accounts for the dynamics of groups of individuals interacting face-to-face.
Our model is able to reproduce many aspects of real-world systems, including the distribution of groups, the correlation in their number, their persistence in time, and the presence of mixing patterns.
Remarkably, the GAM captures characteristics of group conversations in a variety of social contexts, from conferences and exhibitions, where interactions can often be spontaneous, to households in a village, where relationships among group members are more personal and intimate.
The superior performance of the GAM compared to pairwise methods marks the need to adopt higher-order models to investigate groups in face-to-face interactions.

Despite its capability to replicate different features of group interactions, others remain beyond reach.
For instance, our model is Markovian, as agents decide whether to interact with others without memory of the previous time steps. 
Face-to-face interactions, instead, are characterized by complex memory effects, with each group having memory of itself and others \cite{gallo2023higher}.
The asymmetric nature of memory can determine preferred temporal directions in group formation and dispersal that cannot be described by our model, as both dynamics are governed by the same mechanism, i.e., group attractiveness.
Moreover, while smaller groups tend to evolve gradually, with one or few members at the time joining/leaving the group, larger groups can have more complex dynamics \cite{cencetti2021temporal,gallo2023higher,iacopini2023temporal} that the Group Attractiveness Model does not take into account.

In empirical systems, the distributions of group duration are broad-tailed, and, in general, smaller groups have broader distributions than larger ones, i.e., smaller groups last longer.
Although our model reproduces this feature, we observed that denser environments lead to narrower contact duration distributions, likely due to larger volumes of group aggregation and disaggregation.
However, as tuning the density allows us to correctly predict the number of groups of different sizes, a trade-off between the group statistics and their temporal duration remains.
In addition, the data do not provide spatial information about the environment in which contacts take place, making it difficult to determine the appropriate value of agent density.
Further modeling efforts should thus aim at investigating how the spatial dimension, the number of groups, the profile and hierarchical organization of the duration distributions relate to one another. 

\rev{Our modeling framework builds on several simplifications}.
First, it assumes that different stages of group dynamics, i.e., joining pre-existing groups, remaining in them, and leaving them, are governed by a single mechanism, namely the choice of agents to interact or walk away based on the average attractiveness of the groups around them.
In principle, these processes could be \textit{i)} governed by different group properties, or \textit{ii)} regulated by the same quantity, e.g., the group attractiveness, in distinct ways.
\rev{Second, attractiveness is here an inherent feature of conversational groups, as determined by the attractiveness of the individuals forming them.
However, it is reasonable to think that the same group can attract agents in different ways based on their individual characteristics.  
While the extension of our model to include gender relates to this aspect, we believe that further studies in this direction are necessary.}
In connection to this, our model does not distinguish for the types of social groups \cite{lickel2000varieties} and contexts, which naturally determine the dynamics of conversations\rev{, and the structure of social networks in general \cite{newman2003social}.}
\rev{Finally, the definition of the group attractiveness is not unique, as other choices are possible. 
The advantage of the definition in \cref{eq:GA} is twofold:
First, it captures, from a qualitative point of view, the experimental evidence that larger groups are on average less attractive and less stable than smaller ones.
Second, it is parsimonious, meaning that it does not require additional parameters rather than the size of a group and the attractiveness of its members.
We believe analyzing other definitions of group attractiveness that align with these principles represents a valuable direction for further work.}

\rev{Similar to \cite{starnini2013modeling,starnini2016model,oliveira2022group}, the GAM allows for group overlap, namely a situation where face-to-face groups can share some of their members at a given time.
This feature can be unrealistic in most social context and has the potential to affect the behavior of the model.
While we verified that the GAM has, in practice, a low level of overlap that is unlikely to affect our findings (see Fig.~S24), this analysis demands for further investigation into group overlap, given the role that this has on different dynamical processes \cite{malizia2025hyperedge,malizia2025disentangling}.
}

\rev{Though there have been a few attempts to give a higher-order definition of homophily recently \cite{veldt2023combinatorial,sarker2024higher,rizi2025homophily}, our understanding of homophily at the group level remains limited.
Our results advance this line of research by shifting the perspective on how to measure homophily in group interactions. 
Instead of quantifying it \textit{a posteriori}, namely based on mixing patterns in the data, we adopted an \textit{a priori} approach, modeling how microscopic interactions are driven by homophilic preferences.
Yet our model also comes with a few limitations.
First, the GAM is currently limited to a single binary attribute, and further work could aim at extending it to multiple and non-binary attributes.
While the non-binary attribute generalization is straightforward, the multi-attribute case requires more attention.
On the one hand, one could consider a set of matrices for each attribute associated with the agents and thus define a decision-making process over these multiple matrices.
Alternatively, one could adopt an intersectional approach, defining a single set of matrices that modulates the probability of agents to interact based on combinations of attributes, e.g., black-woman, white-man.
Second, we assumed that agents differ only in terms of their intention to interact with their close neighbors, while other factors can be at play, both at an individual and at an attribute level (e.g., one can assume that agent attractiveness correlates with their attributes).
Therefore, one has to be aware that our findings on homophily holds valid under this assumption on the behavior.
Further generalizations of the model including multiple attributes or alternative behavioral processes could provide a robustness check on our findings.
More in general, given the prominence that both group interactions and homophily have (separately) in social systems, a deeper understanding of higher-order homophily is essential.
}

\rev{Beyond ABMs, statistical approaches like the Relational HyperEvent Model (RHEM) \cite{lerner2021dynamic,lerner2023relational} can provide insights on how higher-order contacts depend on group characteristics.
Their focus is on measuring how specific covariates, depending on the characteristics of the members of a group and their previous interactions, affect the group interaction rate.
These empirical methods can serve as a natural complement to our theoretical model: 
While our approach allows us to study whether specific mechanisms (e.g., the group attractiveness) are sufficient to produce observed patterns and generate synthetic data, they enable hypothesis testing on what mechanisms drive group interactions.
}

Overall, our work contributes to the study of human face-to-face interactions through the lens of group dynamics and higher-order mechanisms.
Given its ability to reproduce different features of the data, we are confident that our model will prove to be beneficial in investigating how groups affect different phenomena, including social contagion, epidemic spreading, and the emergence of mixing patterns and segregation in networked populations.

\section*{Methods}
\subsection*{Reconstructing groups from face-to-face pairwise data}
To assess the features of the Group Attractiveness Model, we use datasets from the SocioPatterns collaboration \cite{cattuto2010dynamics}.
These datasets store face-to-face interactions as a list of dyadic contacts with a resolution of 20 seconds.
Therefore, they do not provide any information on group interactions, i.e., face-to-face conversations involving two or more individuals, in the social systems.
However, given the fine-grained temporal resolution of the data, we are able to reconstruct groups of more than two individuals.
Specifically, if at time $t$ in the dataset there are all possible dyadic contacts among $s$ individuals, we can reasonably assume that they are interacting together in a group.
For example, if at time $t$ an individual $i$ is in contact with individuals $j$ and $k$, and these two are also interacting, we can safely say that $i$, $j$ and $k$ form a group of three individuals.

\subsection*{Fitting the homophily matrices}
The Group Attractiveness Model can be extended to assess higher-order mixing patterns in face-to-face interactions.
Specifically, we can enrich the model by tuning the probability that an agent interacts with a neighboring group based on their attributes.
These probabilities are determined by a set of homophily matrices, $H^{(2)}$, $H^{(3)}$, and so on, one for each group size.
Here, we show how we can analytically derive the homophily matrices in the case of a single, binary attribute $\alpha\in\{0,1\}$.
Since larger groups are less abundant in the data, we focus on groups of size two and three, tuned by the matrices $H^{(2)} = [[h_{00}, h_{01}],[h_{10},h_{11}]]$ and $H^{(3)} = [[h_{000}, h_{001}, h_{011}],[h_{100},h_{101},h_{111}]]$ (the superscripts are omitted for simplicity).
$h_{\alpha\beta}$ denotes the probability that an agent with attribute $\alpha$ starts to interact with an agent having attribute $\beta$, while $h_{\alpha\beta\gamma}$ represents the probability that an agent with attribute $\alpha$ starts to interact with a group of two agents having attributes $\beta$ and $\gamma$, respectively.

Groups of two agents can be in three different configurations, namely $(0,0)$, $(0,1)$, and $(1,1)$.
Let us consider the scenario in which two agents that are not interacting, i.e., they are not part of any common group, get in contact and form a group.
We denote the number of pairs in each configuration generated at time $t$ as $E_{00}$, $E_{01}$, and $E_{11}$, respectively.
In general, we can write the number of pairs in configuration $(\alpha,\beta)$ as
\begin{equation}
    E_{\alpha\beta} = G^{(2)}p_{ij,\alpha\beta}.
\end{equation}
$G^{(2)}$ denotes the average number of interactions between two agents (previously not interacting) that could be formed without considering homophily.
$G^{(2)}$ depends on number of agents $N$, their radius of action $d$, the size of the environment $L$, and the attractiveness distribution.
$p_{ij,\alpha\beta}$ represents the probability that two agents $i$ and $j$ (i) have attributes $\alpha$ and $\beta$, respectively, and (ii) start interacting according to their attributes.
A pair is created in three different situations, depending on whether (1) only $i$, (2) only $j$, or (3) both $i$ and $j$ initiate the formation of the group.
Therefore, we can write $p_{ij,\alpha\beta}$ as
\begin{equation}
    p_{ij,\alpha\beta} = p_{i\rightarrow j,\alpha\beta} + p_{i\leftarrow j,\alpha\beta} + p_{i\leftrightarrow j,\alpha\beta}, 
\end{equation}
where the arrows indicate the three possible scenarios of group formation.
In the case of two agents with attributes $\alpha=\beta=0$, we have
\begin{eqnarray}
    p_{ij,00} &&= p_{i\rightarrow j,00} + p_{i\leftarrow j,00} + p_{i\leftrightarrow j,00} \nonumber\\ 
    &&=f_0^2 \times h_{00}(1-h_{00}) + f_0^2 \times (1-h_{00})h_{00} + f_0^2 \times h_{00}^2 \nonumber\\
    &&= f_0^2(1-h_{01}^2),   
\end{eqnarray}
where $f_0$ is the fraction of agents with attribute 0, and we assumed $h_{00} + h_{01} =1$ \cite{oliveira2022group}.
Hence, the number of pairs in state $(0,0)$ generated at time $t$ is
\begin{equation}
    E_{00} = G^{(2)}f_0^2 (1-h_{01}^2).
\end{equation}
Similarly, we can write the number of groups in state $(1,1)$ as
\begin{equation}
    E_{11} = G^{(2)}f_1^2 (1-h_{10}^2),
\end{equation}
where $f_1$ is the fraction of agents with attribute 1, and we assumed $h_{10} + h_{11} =1$.
Finally, the number of groups in state $(0,1)$ is given by
\begin{equation}
    E_{01} = 2 G^{(2)}f_0f_1 (1-h_{00}h_{11}),
\end{equation}
where the factor two comes from the fact that $i$ and $j$ can have either attributes 0 or 1.
We can then normalize the number of groups in each configuration by the total number of groups generated, obtaining the fractions
\begin{eqnarray}
    e_{00} = \frac{f_0^2 (1-h_{01}^2)}{f_0^2 (1-h_{01}^2) + 2f_0f_1 (1-h_{00}h_{11}) + f_1^2 (1-h_{10}^2)}, \nonumber \\
    e_{01} = \frac{2f_0f_1 (1-h_{00}h_{11})}{f_0^2 (1-h_{01}^2) + 2f_0f_1 (1-h_{00}h_{11}) + f_1^2 (1-h_{10}^2)}, \nonumber \\
    e_{11} = \frac{f_1^2 (1-h_{10}^2)}{f_0^2 (1-h_{01}^2) + 2f_0f_1 (1-h_{00}h_{11}) + f_1^2 (1-h_{10}^2)}.
    \label{eq:fractions_two}
\end{eqnarray}
In the case of gender homophily, setting $f_0$ and $f_1$ equal to the fraction of female and male individuals, and $e_{\alpha\beta}$ equal to the average fractions of pairs formed at time $t$, where the individuals were not interacting at time $t-1$, we can estimate the entries of $H^{(2)}$.

If $h_{\alpha\alpha} > 1/2$, agents prefer to interact with those having the same attribute, namely, the system is in a homophilic regime.
Instead, when $h_{\alpha\alpha} < 1/2$ agents tend to interact more with those having the other attribute, i.e., heterophilic regime.
The case $h_{\alpha\alpha} = 1/2$ corresponds to the neutral scenario where agents interact without any preferences.

We now consider the scenario in which an agent with attribute $\alpha$ joins a pair of interacting agents that are within its scope.
We denote the number of groups of size three in configuration $(\alpha,\beta,\gamma)$ generated as $T_{\alpha\rightarrow(\beta,\gamma)}$.
This can be written as
\begin{equation}
    T_{\alpha\rightarrow (\beta,\gamma)} = M^{(2)}\varepsilon_{\beta\gamma}h_{\alpha\beta\gamma},
\end{equation}
where $M^{(2)}$ is the average number of groups of size two within the scope of an agent, $\varepsilon_{\beta\gamma}$ is the fraction of groups in state $(\beta,\gamma)$, while $h_{\alpha\beta\gamma}$ is the element of the homophily matrix $H^{(3)}$ denoting the probability that an agent with attribute $\alpha$ interacts with a pairs of agents with attributes $\beta$ and $\gamma$, respectively.
\rev{Note that $e_{\beta\gamma}$ denotes the fraction of dyads in state $(\beta,\gamma)$ formed at a given time, and differs from $\varepsilon_{\beta\gamma}$, which is the fraction of dyads active at a given time.
This also includes the dyads formed previously that persisted over time, and those generated from other dynamics, e.g., a group of three that loses a member.}
Focusing on $\alpha=0$, we can write the number of groups of size three formed at time $t$ as
\begin{eqnarray}
    T_{0\rightarrow (0,0)} = M^{(2)}\varepsilon_{00}h_{000}, \nonumber \\
    T_{0\rightarrow (0,1)} = M^{(2)}\varepsilon_{01}h_{001}, \nonumber \\
    T_{0\rightarrow (1,1)} = M^{(2)}\varepsilon_{11}h_{011}.
\end{eqnarray}
Normalizing by the total number of groups generated, we obtain the fractions
\begin{eqnarray}
    \tau_{0\rightarrow (0,0)} = \frac{\varepsilon_{00}h_{000}}{\varepsilon_{00}h_{000} + \varepsilon_{01}h_{001} + \varepsilon_{11}h_{011}}, \nonumber \\
    \tau_{0\rightarrow (0,1)} = \frac{\varepsilon_{01}h_{001}}{\varepsilon_{00}h_{000} + \varepsilon_{01}h_{001} + \varepsilon_{11}h_{011}}, \nonumber \\
    \tau_{0\rightarrow (1,1)} = \frac{\varepsilon_{11}h_{011}}{\varepsilon_{00}h_{000} + \varepsilon_{01}h_{001} + \varepsilon_{11}h_{011}}.
    \label{eq:fractions_three_0}
\end{eqnarray}
Similarly, for an agent with attribute 1, we find
\begin{eqnarray}
    \tau_{1\rightarrow (0,0)} = \frac{\varepsilon_{00}h_{100}}{\varepsilon_{00}h_{100} + \varepsilon_{01}h_{101} + \varepsilon_{11}h_{111}}, \nonumber \\
    \tau_{1\rightarrow (0,1)} = \frac{\varepsilon_{01}h_{101}}{\varepsilon_{00}h_{100} + \varepsilon_{01}h_{101} + \varepsilon_{11}h_{111}}, \nonumber \\
    \tau_{1\rightarrow (1,1)} = \frac{\varepsilon_{11}h_{111}}{\varepsilon_{00}h_{100} + \varepsilon_{01}h_{101} + \varepsilon_{11}h_{111}}.
    \label{eq:fractions_three_1}
\end{eqnarray}
Assuming that $h_{000} + h_{001} + h_{011} = h_{100} + h_{101} + h_{111} = 1$, we can estimate the entries of the homophily matrix $H^{(3)}$.
Specifically, we set $\varepsilon_{\beta\gamma}$ equal to the fractions of unique groups of size two in the different gender configurations, while $\tau_{\alpha\rightarrow (\beta,\gamma)}$ can be evaluated by counting how many times in the data a pair of interacting individuals at time $t-1$ is followed by a group of size three at time $t$.
Note that the neutral scenario with no homophilic preferences in the group formation corresponds to $h_{000} = h_{011} = 1/3$ and $h_{100} = h_{111} = 1/3$.

Finally, we can recover the Group Attractiveness Model without homophily by assuming that all agents have the same attribute, say $f_0 = 1$, and that the corresponding homophilic interaction probabilities are equal to 1, namely $h_{00} = 1$ (no matter the value of $h_{11}$) and $h_{000} = 1$ (no matter the values of $h_{100}$ and $h_{111}$).

\rev{
We can estimate the other homophily matrices following the same derivation applied to $H^{(3)}$.
To derive $H^{(d)}$, one needs to estimate $2d$ elements from $2d$ equations describing how agents join groups of size $d-1$, together with 2 constraints on the row sum of the matrix.
Formally, one has to solve the system of equations
\begin{equation}
    \begin{split}
        &\tau_{\nu\rightarrow \sigma^{(d-1)}} = \frac{f\left(\sigma^{(d-1)}\right)h_{\nu\sigma^{(d-1)}}}{\sum\limits_{\sigma^{(d-1)}}{f\left(\sigma^{(d-1)}\right)h_{\nu\sigma_{d-1}}}} \\
        & \sum\limits_{\sigma^{(d-1)}}h_{\nu\sigma^{(d-1)}} = 1
    \end{split},
\end{equation}
where $\nu\in[0,1]$, $\sigma^{(d-1)}$ is one of the $d$ possible states for groups of size $d-1$, $\tau_{\nu\rightarrow \sigma^{(d-1)}}$ is the fraction of transitions in which agents with attribute $\nu$ join groups in state $\sigma^{(d-1)}$, $f(\sigma^{(d-1)})$ is the fraction of groups of size $d-1$ in state $\sigma^{(d-1)}$, while $h_{\nu\sigma^{(d-1)}}$ is an element of the homophily matrix $H^{(d)}$.
Note that each element of $H^{(d)}$ has $d$ indices, one being $\nu$, and the other $d-1$ coming from $\sigma^{(d-1)}$.
As done for $d=3$ with $\tau_{\nu\rightarrow (\beta,\gamma)}$ and $\varepsilon_{\beta\gamma}$, one can estimate $\tau_{\nu\rightarrow \sigma^{(d-1)}}$ and $f(\sigma^{(d-1)})$ from the data, to finally derive $H^{(d)}$.
}

\section*{Data Availability}
Data on contacts in scientific conferences are available upon request at https://doi.org/10.7802/235.
All other datasets are freely available at https://www.sociopatterns.org/datasets.

\section*{Code Availability}
A Python implementation of the Group Attractiveness Model is available as part of the HGX library \cite{lotito2023hypergraphx}.

\bibliography{biblio}

\section*{Acknowledgments}
L.G. thanks Rahel Geppert and Rebeka O. Szabo for the insightful suggestions and discussion.
L.G. acknowledges support from the Villum Foundation (project no. 57396) at the University of Copenhagen.
L.G. and F.B. acknowledge support of the Air Force Office of Scientific Research under award number FA8655-22-1-7025. 
C.Z. acknowledges funding by the European Union under Horizon EU project LearnData, 101086712.
C.Z. acknowledges support from the Villum Foundation (project no. 37394) at the University of Copenhagen.
\end{document}

\title{Supplementary Information of the manuscript \\ ``Higher-order modeling of face-to-face interactions''}% Force line breaks with \\

\author{Luca Gallo}
    \affiliation{
        Department of Network and Data Science, Central European University, 1100 Vienna, Austria
    }
    \affiliation{
        ANETI Lab, Corvinus Institute for Advanced Studies (CIAS), Corvinus University, 1093, Budapest, Hungary
    }
    \affiliation{
        Center for Social Data Science (SODAS), University of Copenhagen, 1353, Copenhagen, Denmark
    }

\author{Chiara Zappal\`a}%
    \affiliation{
        Center for Social Data Science (SODAS), University of Copenhagen, 1353, Copenhagen, Denmark
    }
    \affiliation{
        Center for Collective Learning, Corvinus Institute for Advanced Studies (CIAS), Corvinus University, 1093 Budapest, Hungary
    }

\author{Fariba Karimi}%
    \affiliation{
        Graz University of Technology, 8010 Graz, Austria
    }
    \affiliation{
        Complexity Science Hub, A-1080 Vienna, Austria
    }

\author{Federico Battiston}%
    \affiliation{
        Department of Network and Data Science, Central European University, 1100 Vienna, Austria
    }

\date{\today}

\maketitle

\section*{Calibrating the models of human face-to-face interactions}
We study ten datasets from the SocioPatterns collaboration \cite{cattuto2010dynamics}.
Those contain information on the dynamics of face-to-face interactions in different social contexts, namely a primary school (``PS'') \cite{stehle2011high}, a high-school (``HS11'', ``HS12'', and ``HS13'')\cite{fournet2014contact,mastrandrea2015contact}, two scientific conferences (``C16'' and ``C17'') \cite{genois2019building}, an art exhibition (``SG'') \cite{isella2011s}, a hospital (``H'') \cite{vanhems2013estimating}, a village in Malawi (``MV'') \cite{ozella2021using}, and a workplace (``WP'') \cite{genois2018can}, recorded using active Radio Frequency Identification (RFID) devices.
\rev{Additionally, we analyze data from experiments conducted in a middle school (``MS'') and an elementary school (``ES'') in Utah \cite{toth2015role}, recorded with Wireless Ranging-Enabled Nodes (WRENs), and data from the Copenhagen Network Study \cite{sapiezynski2019interaction}, consisting of contacts in a university campus (``UC'') registered every using the Bluetooth technology. 
Here we focus on the six SocioPatterns datasets with gender, and report the analysis of the other six systems in the Supplementary Information.
}

All datasets store contacts as pairwise interactions, so we first reconstruct group interactions among individuals leveraging the fine-grained temporal information of the data (see Methods in the main text for details).
Then, we construct the hypergraph \cite{battiston2020networks} of face-to-face interactions in the system.
In particular, we represent each individual as a node and each group interaction as a hyperedge connecting the nodes representing the members of the group.
Next, we identify the largest connected component of such hypergraph, and discard the nodes that are not part of it as well as the groups they take part in. 
The number of individuals in each social system, $N$, and the number of groups of different sizes are reported in \cref{tab:S1}.

\newcolumntype{C}{>{\centering\arraybackslash}p{3.5em}}
\begin{table}[h!]
    \centering
    \begin{tabular}{ccCCCCCCCCCCcc}
         \multirow{2}{*}{Dataset} & Number & \multicolumn{10}{c}{Number of groups of size} & \multirow{2}{*}{$L_\mathrm{GAM}$} & \multirow{2}{*}{$L_\mathrm{AM}$} \\
         \cline{3-12}
          & of agents & 2 & 3 & 4 & 5 & 6 & 7 & 8 & 9 & 10 & 11 &  &  \\
          \hline
         HS11 & 126 & 1608 & 470 & 50 & 1 & - & - & - & - & - & - & 20.5 & 34 \\
         HS12 & 180 & 2125 & 466 & 50 & 4 & - & - & - & - & - & - & 29 & 46 \\
         HS13 & 320 & 5277 & 1990 & 211 & 7 & - & - & - & - & - & - & 28 & 50 \\
         PS & 227 & 7069 & 4094 & 322 & 9 & - & - & - & - & - & - & 19.5 & 36 \\
         C16 & 115 & 4875 & 6410 & 1611 & 205 & 19 & 1 & - & - & - & - & 10.6 & 19 \\
         C17 & 207 & 11036 & 4969 & 618 & 45 & 2 & - & - & - & - & - & 25.5 & 43 \\
         SG & 279 & 1398 & 410 & 43 & 3 & - & - & - & - & - & - & 27 & 45 \\
         H & 75 & 1108 & 657 & 58 & 2 & - & - & - & - & - & - & 10.6 & 19.5 \\
         MV & 86 & 342 & 86 & 4 & - & - & - & - & - & - & - & 15.7 & 26.5 \\
         WP & 92 & 742 & 44 & 2 & - & - & - & - & - & - & - & 48 & 63 \\
         \rev{MS} & \rev{591} & \rev{49301} & \rev{23850} & \rev{2500} & \rev{241} & \rev{28} & \rev{16} & \rev{4} & \rev{5} & \rev{2} & \rev{-} & \rev{38} & \rev{67} \\
         \rev{ES} & \rev{339} & \rev{14184} & \rev{10365} & \rev{1683} & \rev{228} & \rev{56} & \rev{22} & \rev{3} & \rev{-} & \rev{-} & \rev{-} & \rev{21} & \rev{35} \\
         \rev{UC} & \rev{692} & \rev{50977} & \rev{113842} & \rev{150622} & \rev{122843} & \rev{76032} & \rev{35976} & \rev{11449} & \rev{1998} & \rev{201} & \rev{14} & \rev{13.4} & \rev{20} \\
    \end{tabular}
    \caption{\textbf{Brief summary of the relevant information for each dataset}. We report the total number of individuals considered and the number of groups of different sizes extracted from the data. In addition, the values of the environment size $L$ that best fit the group statistics for the Group Attractiveness Model (GAM) and the Attractiveness Model (AM) are shown.}
    \label{tab:S1}
\end{table}

In the Group Attractiveness Model, as in the Attractiveness Model \cite{starnini2013modeling}, the number of groups formed depends to a great extent on the density of agents.
When the size $L$ of the environment is significantly large, the agents rarely get in contact with each other, and the formation of large groups becomes unlikely. 
On the other hand, when $L$ is small, the density is high and the agents are close to each other, so it is possible to observe a wider variety of group sizes.
Therefore, for both models, we estimate the value of $L$ that best fits the distribution of groups in the datasets.
Specifically, we run 100 simulations, stopping each repetition once the number of groups of size two generated matches the empirical value.
We evaluate the average number of groups of different sizes generated over the different repetitions and estimate how different the model prediction is from the data using the mean squared error, namely
\begin{equation}
    E = \frac{1}{\mathrm{S}-1}\sum\limits_{s=2}^{\mathrm{S}}\left[M_s - \widehat{M}_s(L)\right]^2
\end{equation}
where $M_s$ denotes the number of groups of size $s$ present in the data, $\widehat{M}_s$ is the average number of groups of size $s$ predicted by the model, and $\mathrm{S}$ is the maximal group size in the data.
We repeat this procedure with various $L$ and select the value that minimizes the calibration error $E$.
The best-fit values of $L$ for both models considered are reported in \cref{tab:S1}.
In general, the value of $L$ estimated for the Group Attractiveness Model is smaller than that obtained with the Attractiveness Model.

{\color{crimson}
\section*{Dependence of the GAM on the distribution of attractiveness}
Interactions in the GAM are mediated by the attractiveness of agents, defined as their power to make others willing to interact with them.
This characteristic is contextual and likely to be influenced by several factors, so it may be hard to measure empirically.
Therefore, we assumed to sample the individual attractiveness from a uniform distribution because (i) it is the most unbiased probability distribution, and (ii) it is the distribution considered in \cite{starnini2013modeling}.

Here, we prove that our model is robust to different choices of attractiveness distributions.
We sample attractiveness from a normal distribution centered on $\langle a\rangle=0.5$ and with $\sigma=0.25$, truncated so that $a\in[0,1]$.
The underlying assumption here is that a typical value of attractiveness exists, meaning that agents have a similar social appeal.

We compare how the uniform and the truncated normal distributions of attractiveness affect the group statistics, correlation and burstiness in the GAM. 
For both attractiveness distributions, we run 100 simulations with $N=300$ agents, in a square environment of size $L=30$.
All other parameters are the same as in the main text.
The simulations end after 5000 iterations or once the number of unique dyadic contacts reaches $30\%$ of all potentials dyads.

\begin{figure}[h]
    \centering
    \includegraphics[width=.5\columnwidth]{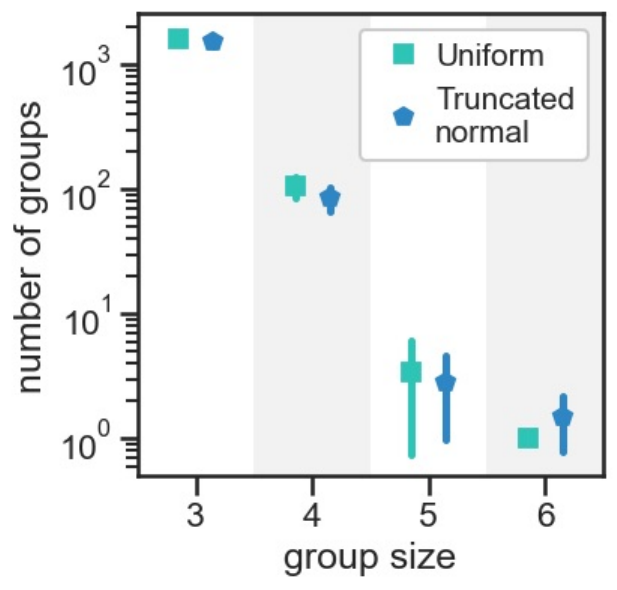}
    \caption{\rev{\textbf{Group statistics as a function of the attractiveness distribution}. 
    Distribution of unique groups of different sizes when attractiveness is sampled from a uniform (teal squares) or a truncated normal (cerulean pentagons).}}
    \label{fig:rev_S01}
\end{figure}

\begin{figure}[h]
    \centering
    \includegraphics[width=.66\columnwidth]{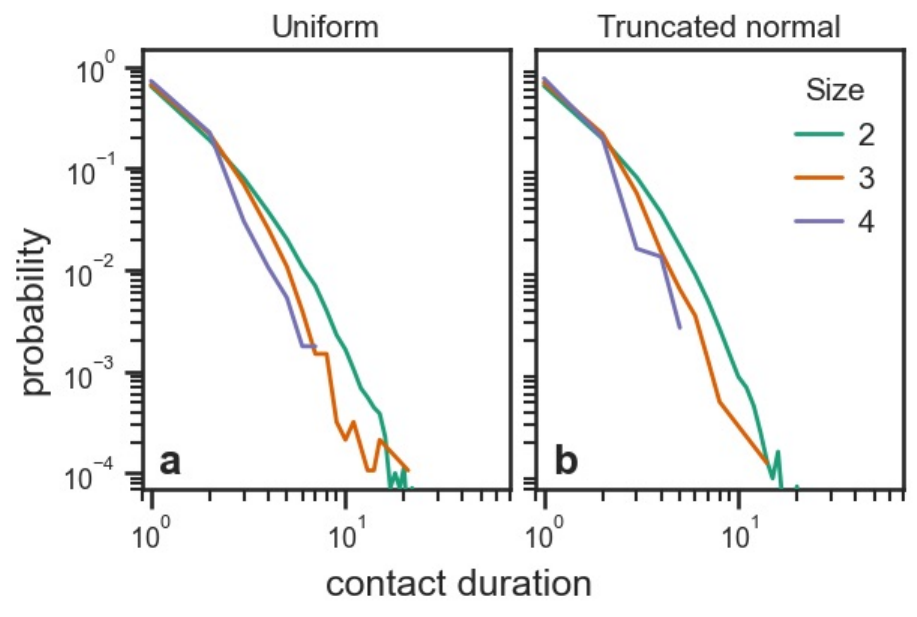}
    \caption{\rev{\textbf{Burstiness as a function of the attractiveness distribution}. 
    Distribution of contact duration for groups of different sizes when attractiveness is sampled from a uniform (Panel \textbf{a}) or a truncated normal (Panel \textbf{b}).}}
    \label{fig:rev_S02}
\end{figure}

Sampling the attractiveness from a truncated normal distribution does not affect the average number of groups of different sizes observed when considering a uniform distribution of attractiveness (see \cref{fig:rev_S01}).
Additionally, we find no differences in the correlation between the number of groups of size 2 and 3 that agents join throughout the simulations ($0.87\pm0.02$ with a uniform distribution of attractiveness, $0.85\pm0.02$ with a truncated normal distributions).
Finally, we observe similar distributions of contact durations as a function of the group size \cref{fig:rev_S02}.
These results show that behavior of GAM can be robust to different assumptions on the distribution of attractiveness.
Future work should consider other assumptions on the distribution of attractiveness, investigating their effects on the collective dynamics of the GAM.
}

\section*{Higher-order statistics of human face-to-face interactions: analysis of additional SocioPatterns datasets}
While in the main text we focus on the six SocioPatterns datasets containing information about participants' gender, here we report the analysis of the other SocioPatterns experiments considered in this study, namely an art exhibition (``SG'') \cite{isella2011s}, a hospital (``H'') \cite{vanhems2013estimating}, a village in Malawi (``MV'') \cite{ozella2021using}, and a workplace (``WP'') \cite{genois2018can}.

\begin{figure}[h]
    \centering
    \includegraphics[width=\columnwidth]{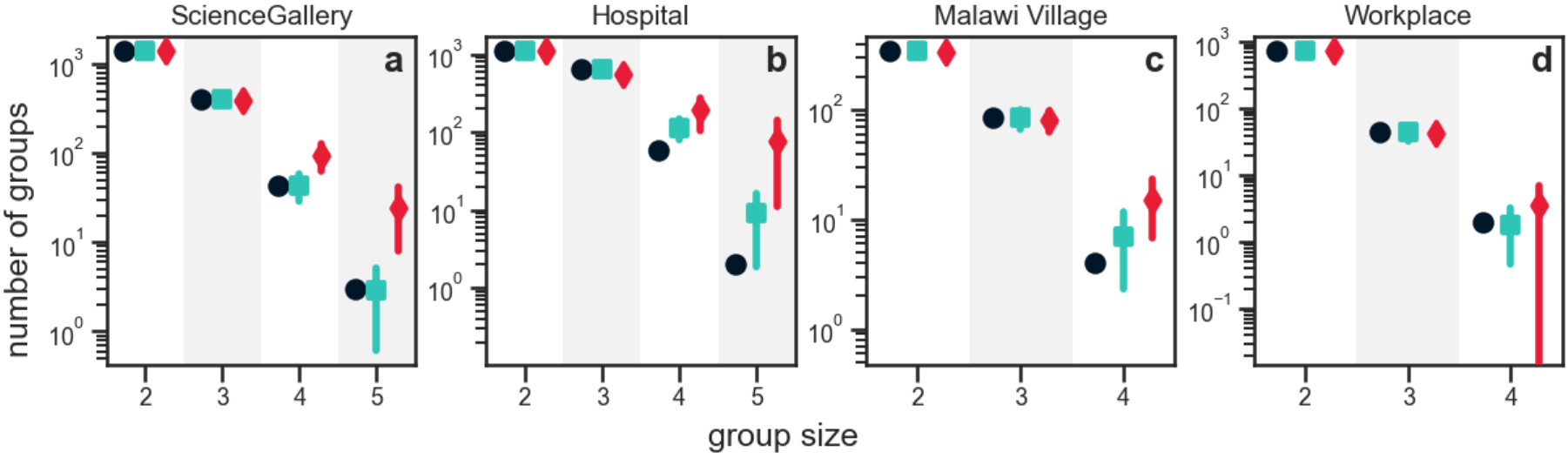}
    \caption{\textbf{The GAM reproduces unique group statistics}. 
    Panels \textbf{a} to \textbf{d} report the distribution of unique groups of different sizes in various social systems (black circles), as well as the predictions of the Group Attractiveness Model (blue squares) and the Attractiveness Model (red diamonds).}
    \label{fig:rev_S1}
\end{figure}

First, we investigate the statistics of unique groups of different sizes. 
We observe that smaller groups are more abundant than larger ones in all systems considered (black circles in \cref{fig:rev_S1}).
The distribution of groups by size is correctly captured by the Group Attractiveness Model (blue squares in \cref{fig:rev_S1}), while the (individual) Attractiveness Model tends to overestimate the number of larger groups.
In general, even in those cases where the Attractiveness Model fits empirical data well, e.g., in the Workplace dataset, the Group Attractiveness model shows better agreement on average.
These additional results corroborate the importance of considering group attractiveness to properly model face-to-face interactions beyond dyadic contacts.

\begin{figure}[p]
    \centering
    \includegraphics[width=.5\columnwidth]{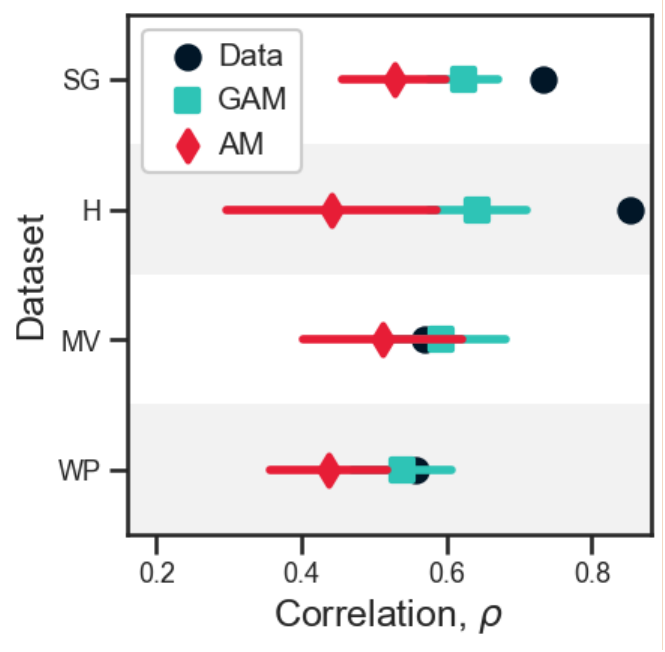}
    \caption{\textbf{The GAM reproduces the correlation between the number of groups of sizes two and three}. Correlation in the number of groups of size two and three in different social contexts (black circles), compared to predictions of the Group Attractiveness Model (blue squares) and the Attractiveness Model (red diamonds).}
    \label{fig:rev_S2}
\end{figure}

Next, we investigate whether individuals participating in groups of a given size also participate in groups of a different size.
In particular, we study the correlations in the number of groups, focusing on groups of sizes two and three.
\cref{fig:rev_S2} shows that face-to-face interactions in pairs and triads are typically highly correlated in real-world systems (black circles), with $\rho$ varying between 0.56 and 0.85.
This is well reproduced by the GAM (blue squares), for which the average correlation coefficient never goes below 0.54.
Furthermore, the GAM can predict the exact value of $\rho$ for half of the systems while slightly underestimating it for the others.
As also shown in the main text, the AM tends to underestimate the correlations between groups of sizes two and three (red diamonds), with values consistently below 0.53, down to 0.33.
Finally, we note that, even in those cases where the GAM is not able to capture the exact value of correlation observed in empirical systems, the GAM shows better agreement with the data than the AM, further pointing out the benefit of considering a higher-order approach to model face-to-face interactions.

\clearpage

{\color{crimson}
\section*{Higher-order statistics of human face-to-face interactions: analysis of alternative sensor technology}
In the main text we focus on SocioPatterns datasets. 
These report dyadic contacts record using Radio Frequency Identification (RFID) devices.
Here, we focus on three further datasets built using different sensing technologies.
Two of them report data on contacts in an elementary school (``ES'') and a middle school (``MS'') in Utah \cite{toth2015role}, collected using wireless ranging enabled nodes (WRENs), while the third one comprises interactions in a university campus (``UC'') in Copenhagen \cite{sapiezynski2019interaction}, recorded via the Bluetooth technology.
Utah´s school data is similar to SocioPatterns': Interactions are recorded every 20 seconds and individual should be proximate (less than 2 meters in the Utah experiments, less than 1.5 meters in the SocioPatterns ones).
However, data from the Copenhagen Network Study (CNS) differs from the others, as the range of the Bluetooth technology is 10 meters and the temporal resolution is of 5 minutes.
To retain only proximate contacts, we filtered out interactions with an associated received signal strength indicator (rssi) smaller than -80 dB.

\begin{figure}[h]
    \centering
    \includegraphics[width=\columnwidth]{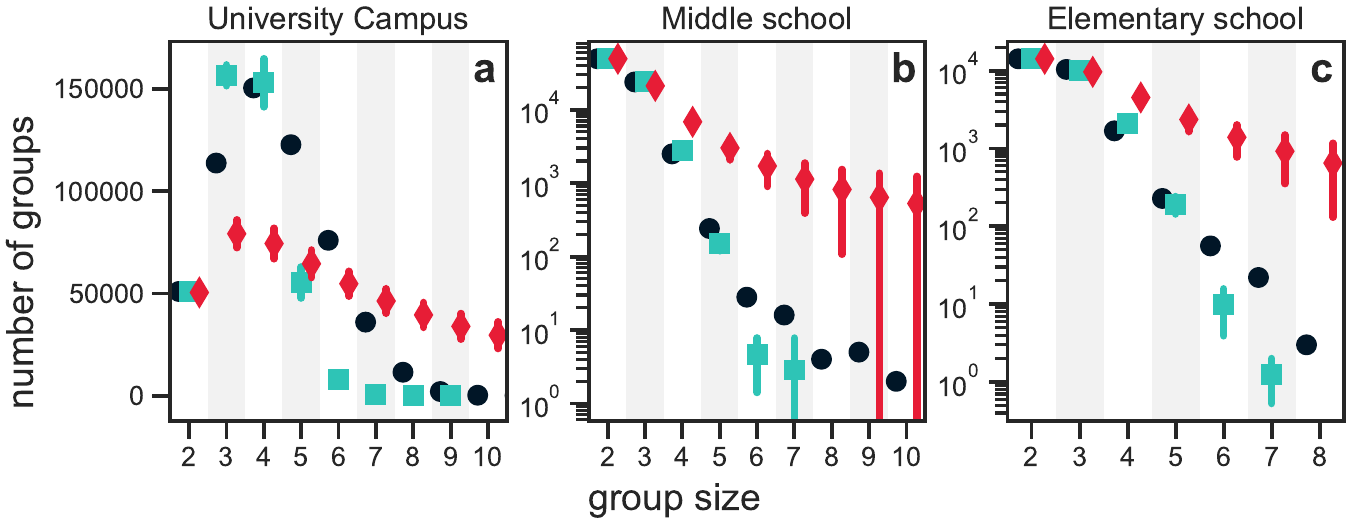}
    \caption{\rev{\textbf{The GAM reproduces unique group statistics in high temporal resolution data}. 
    Panels \textbf{a} to \textbf{c} report the distribution of unique groups of different sizes in various social systems (black circles), as well as the predictions of the Group Attractiveness Model (blue squares) and the Attractiveness Model (red diamonds).}}
    \label{fig:rev_S1_add}
\end{figure}

In the Utah data, smaller groups are more abundant than larger ones, similarly to SocioPatterns data (black circles in \cref{fig:rev_S1_add}).
However, the CNS data show notable differences, as groups of sizes three to six are more abundant than dyads.
The distributions generated by the Group Attractiveness Model (blue squares in \cref{fig:rev_S1_add}) are in good agreement with the Middle school and Elementary school data, while we observe significant differences with the University campus data.
The Attractiveness Model (red diamonds in \cref{fig:rev_S1_add}), instead, fails to reproduce all distributions.

This finding may result from how we inferred group interactions from dyadic contacts.
The  method of promoting the maximal cliques in each temporal snapshots to hyperedges (see Methods) relies on the hypothesis that cliques are more likely the dyadic representation of non-dyadic (group) interactions than a collection of dyadic interactions. 
This hypothesis is reliable for the SocioPatterns and Utah school data, as temporal snapshots represent interactions occurring in a 20-second interval.
Yet the hypothesis is weak when data are less temporally resolved, like in the university campus data, which has a 5-minute resolution.
In this interval, groups may undergo multiple dynamical process (see also \cite{gallo2023higher}).
However, as the whole dynamics is projected onto a single static network, this may lead to generally larger cliques.
This aligns with the observation that the GAM seems to capture the shape of the distribution but underestimates the average group size.

\begin{figure}[h]
    \centering
    \includegraphics[width=.5\columnwidth]{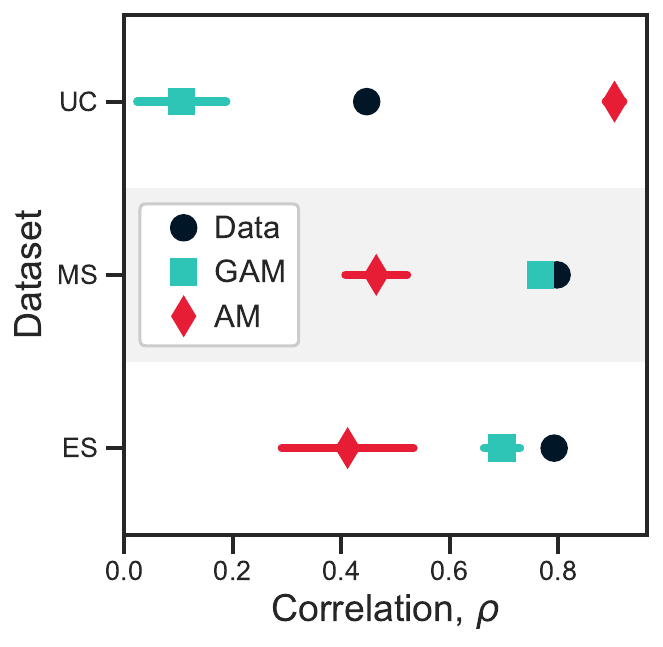}
    \caption{\rev{\textbf{The GAM reproduces the correlation between the number of groups of sizes two and three in high-resolution data}. Correlation in the number of groups of size two and three in different social contexts (black circles), compared to predictions of the Group Attractiveness Model (blue squares) and the Attractiveness Model (red diamonds).}}
    \label{fig:rev_S2_add}
\end{figure}

The analysis of the correlations in the number of groups also aligns with this interpenetration. 
The correlation between the number of groups of size two and three is large in the Utah data (0.80 in the ``MS'' data, and 0.79 in the ``ES'' data, well captured by the GAM), while it drops to 0.45 in the CNS data.
This is substantially different from all other datasets, where the value of correlation almost never goes below 0.7 (see Results and the previous section of this Supplementary Information).
This finding may also be explained by the lower temporal resolution of the CNS data:
As large cliques may form within 5-minutes intervals, groups of size two (as defined in our inference method) become less likely, so its number starts anti-correlating with the number of larger groups, including those of size 3. 
This is similar to what happens in the Attractiveness Model, which generally has a low value of correlation (see again Results).

Overall, these findings show that the GAM is robust to different sensing technologies, i.e., the WRENs.
They also highlight the limitation of the group inference approach in dataset without a high temporal resolution, and the need for alternative methodologies to extract groups from dyadic interactions.  
}

\section*{Burstiness in higher-order interactions}
In human face-to-face interactions, the duration of contacts between individuals displays broad-tailed distributions, indicating that most contacts are brief and few last for long periods of time, with no characteristic scale.
The distributions of contact duration are typically organized in a hierarchy: Small groups show broader distributions compared to larger ones.
Whereas the Attractiveness Model fails to capture this hierarchical organization of group burstiness, the Group Attractiveness Model correctly predicts it.
In the main text, we focus on the ``HS11'' dataset; here, we show that the emergence of a hierarchical structure is a common feature of different social systems that our model is able to replicate.

\begin{figure}[hb!]
    \centering
    \includegraphics[width=\columnwidth]{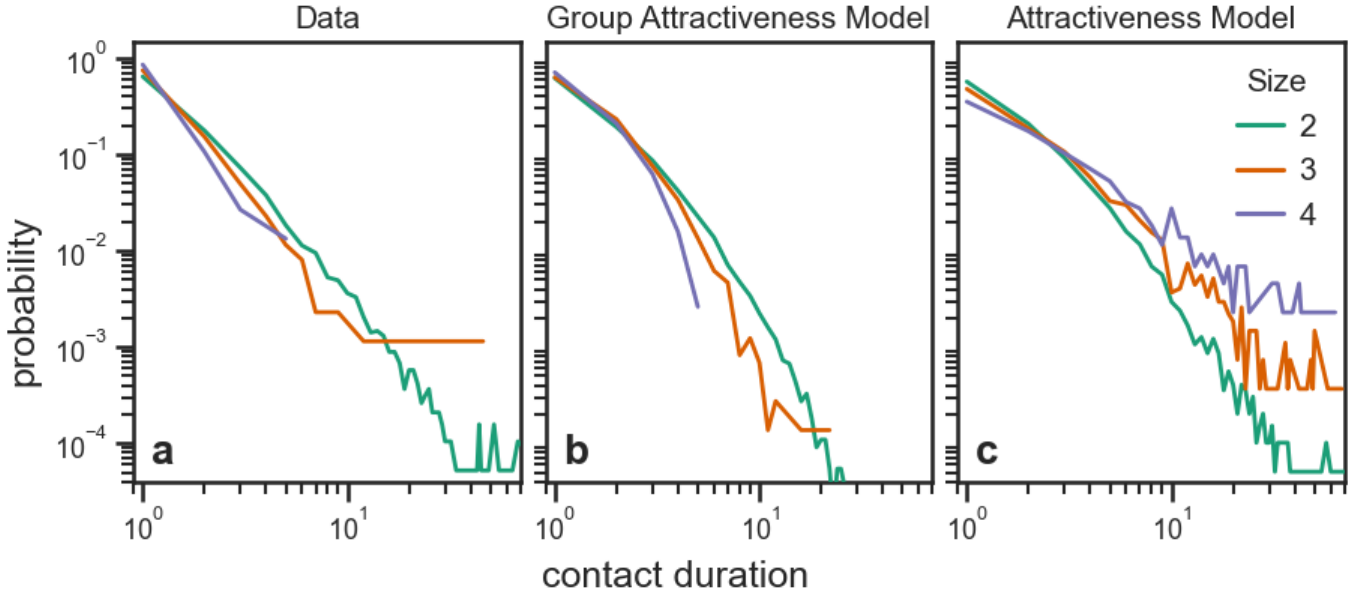}
    \caption{\textbf{Burstiness in the High-school 2012 dataset}. We show the distributions of contact duration for groups of different sizes in the ``HS12'' dataset (panel \textbf{a}), as well as those predicted by the Group Attractiveness Model (panel \textbf{b}) and the Attractiveness Model (panel \textbf{c}).}
    \label{fig:S1}
\end{figure}

\begin{figure}[hb!]
    \centering
    \includegraphics[width=\columnwidth]{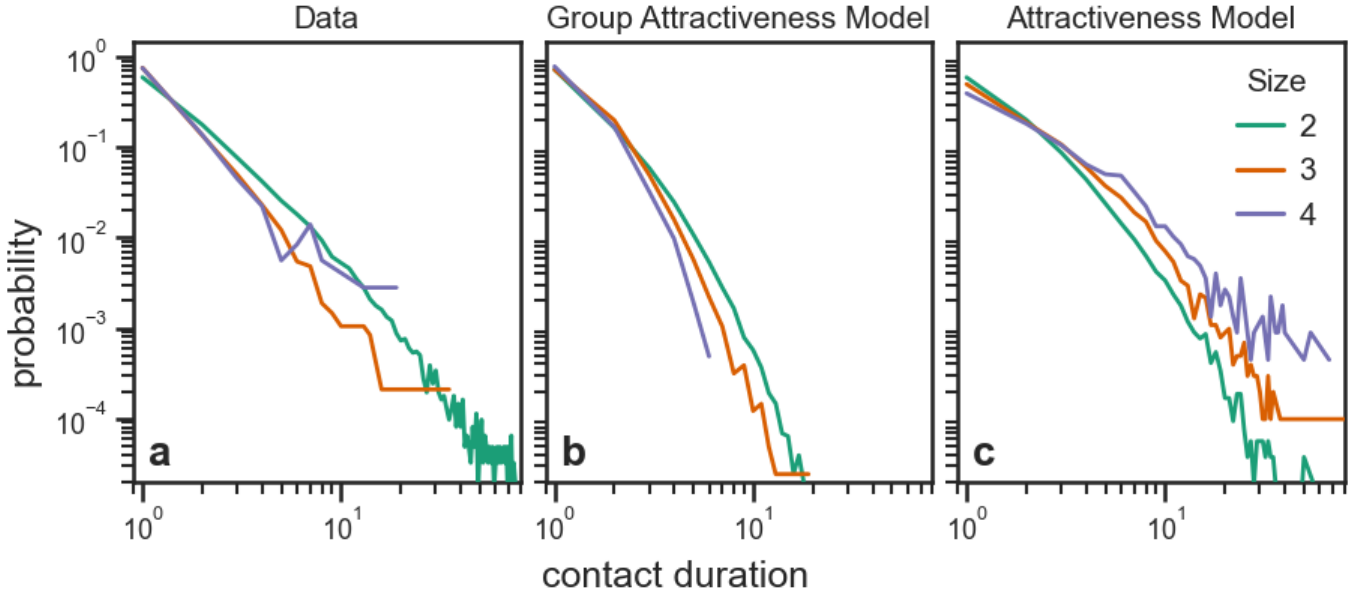}
    \caption{\textbf{Burstiness in the High-school 2013 dataset}. We show the distributions of contact duration for groups of different sizes in the ``HS13'' dataset (panel \textbf{a}), as well as those predicted by the Group Attractiveness Model (panel \textbf{b}) and the Attractiveness Model (panel \textbf{c}).}
    \label{fig:S2}
\end{figure}

\begin{figure}[p]
    \centering
    \includegraphics[width=\columnwidth]{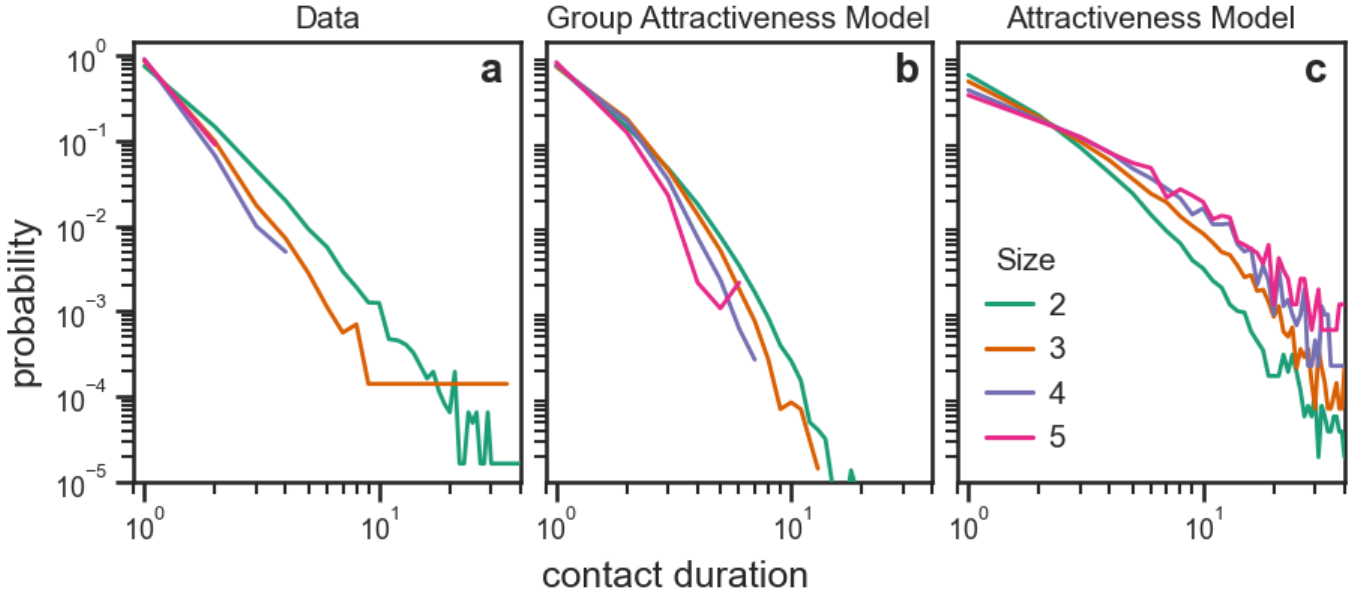}
    \caption{\textbf{Burstiness in the Primary school dataset}. We show the distributions of contact duration for groups of different sizes in the ``PS'' dataset (panel \textbf{a}), as well as those predicted by the Group Attractiveness Model (panel \textbf{b}) and the Attractiveness Model (panel \textbf{c}).}
    \label{fig:S3}
\end{figure}

\begin{figure}[p]
    \centering
    \includegraphics[width=\columnwidth]{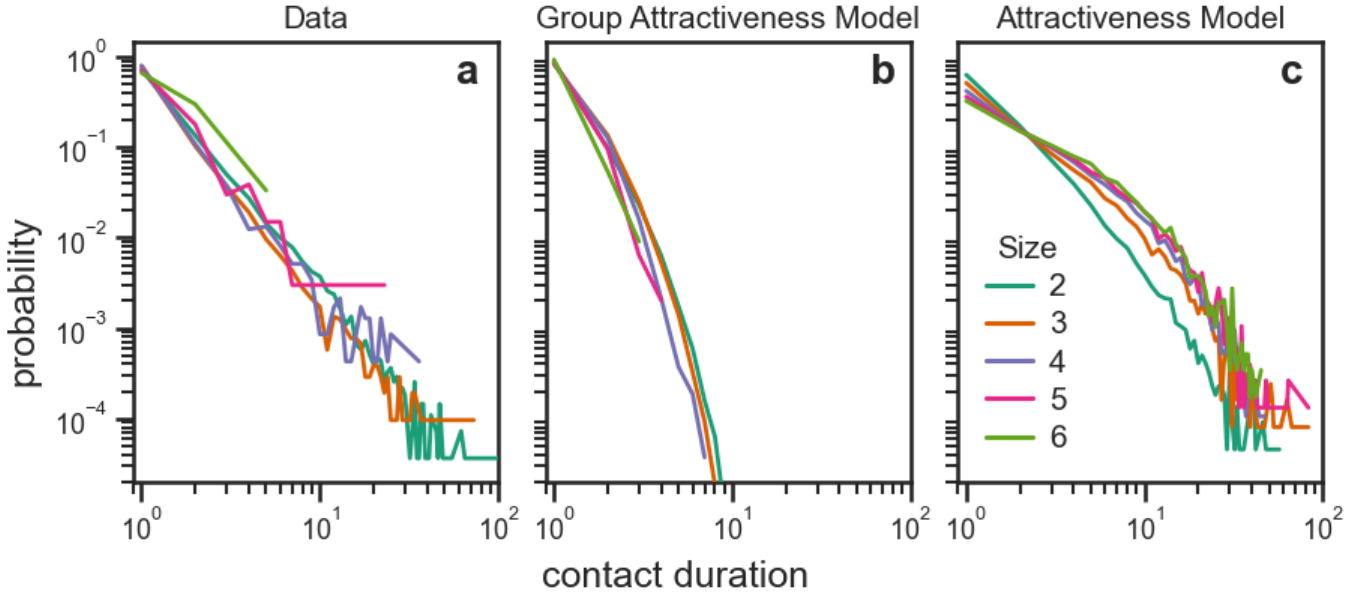}
    \caption{\textbf{Burstiness in the Conference 2016 dataset}. We show the distributions of contact duration for groups of different sizes in the ``C16'' dataset (panel \textbf{a}), as well as those predicted by the Group Attractiveness Model (panel \textbf{b}) and the Attractiveness Model (panel \textbf{c}).}
    \label{fig:S4}
\end{figure}

\begin{figure}[p]
    \centering
    \includegraphics[width=\columnwidth]{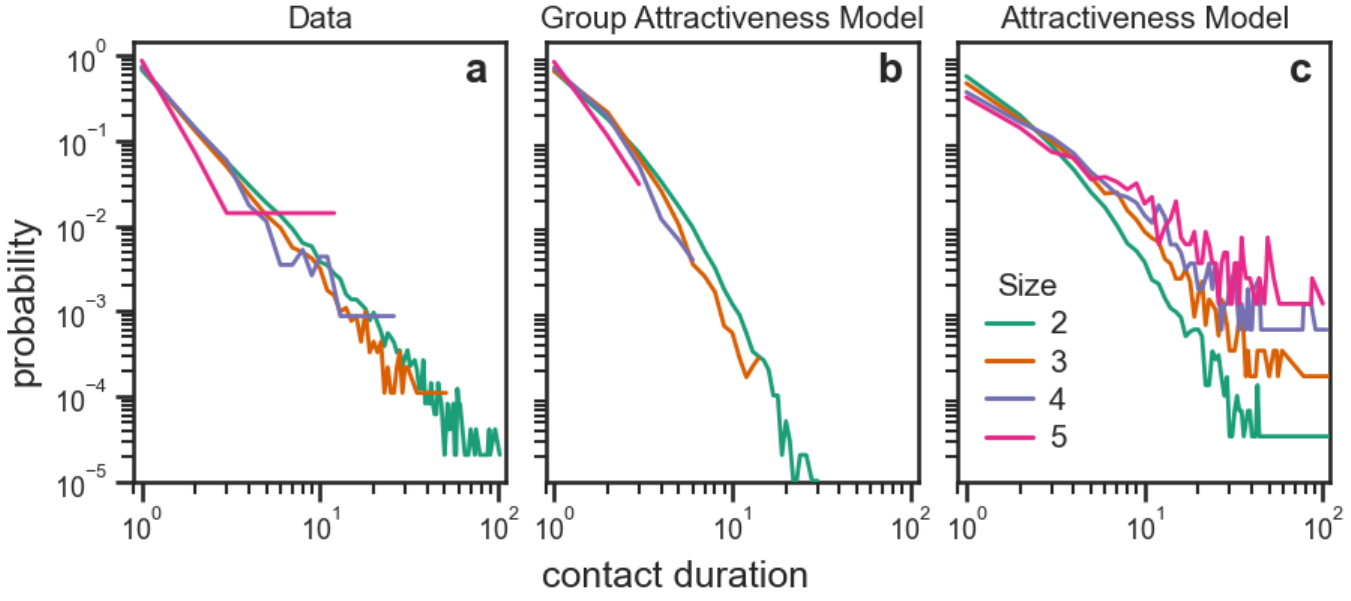}
    \caption{\textbf{Burstiness in the Conference 2017 dataset}. We show the distributions of contact duration for groups of different sizes in the ``C17'' dataset (panel \textbf{a}), as well as those predicted by the Group Attractiveness Model (panel \textbf{b}) and the Attractiveness Model (panel \textbf{c}).}
    \label{fig:S5}
\end{figure}

\begin{figure}[p]
    \centering
    \includegraphics[width=\columnwidth]{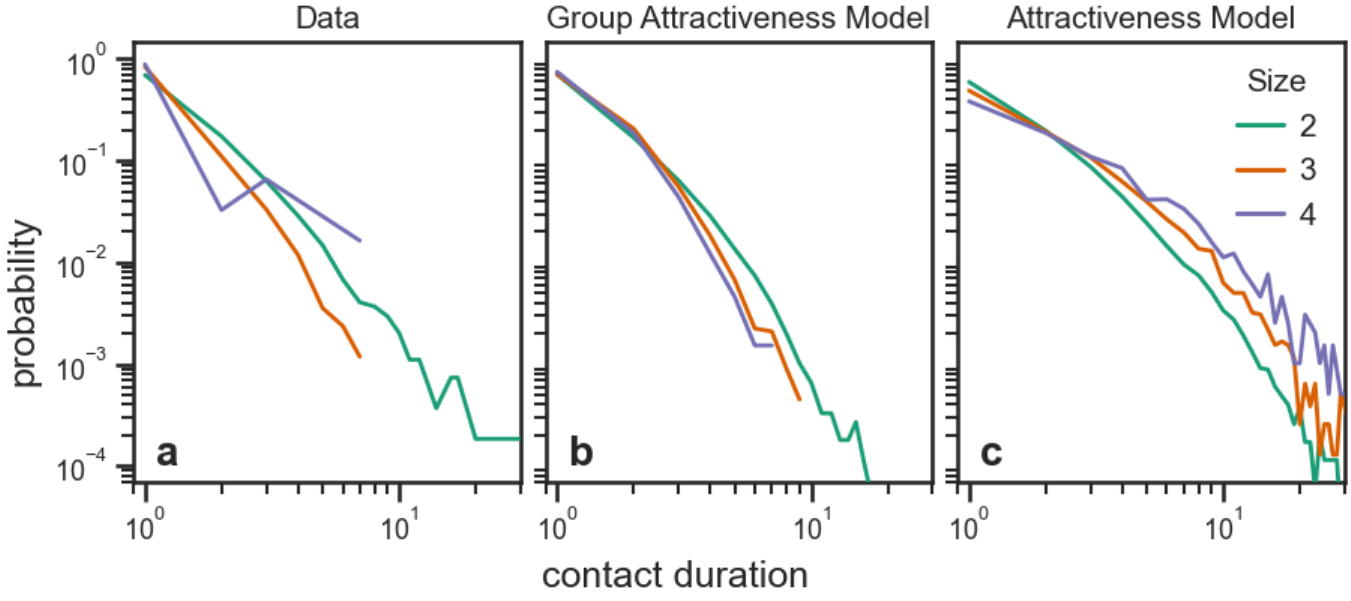}
    \caption{\textbf{Burstiness in the ScienceGallery dataset}. We show the distributions of contact duration for groups of different sizes in the ``SG'' dataset (panel \textbf{a}), as well as those predicted by the Group Attractiveness Model (panel \textbf{b}) and the Attractiveness Model (panel \textbf{c}).}
    \label{fig:rev_S3}
\end{figure}

\begin{figure}[p]
    \centering
    \includegraphics[width=\columnwidth]{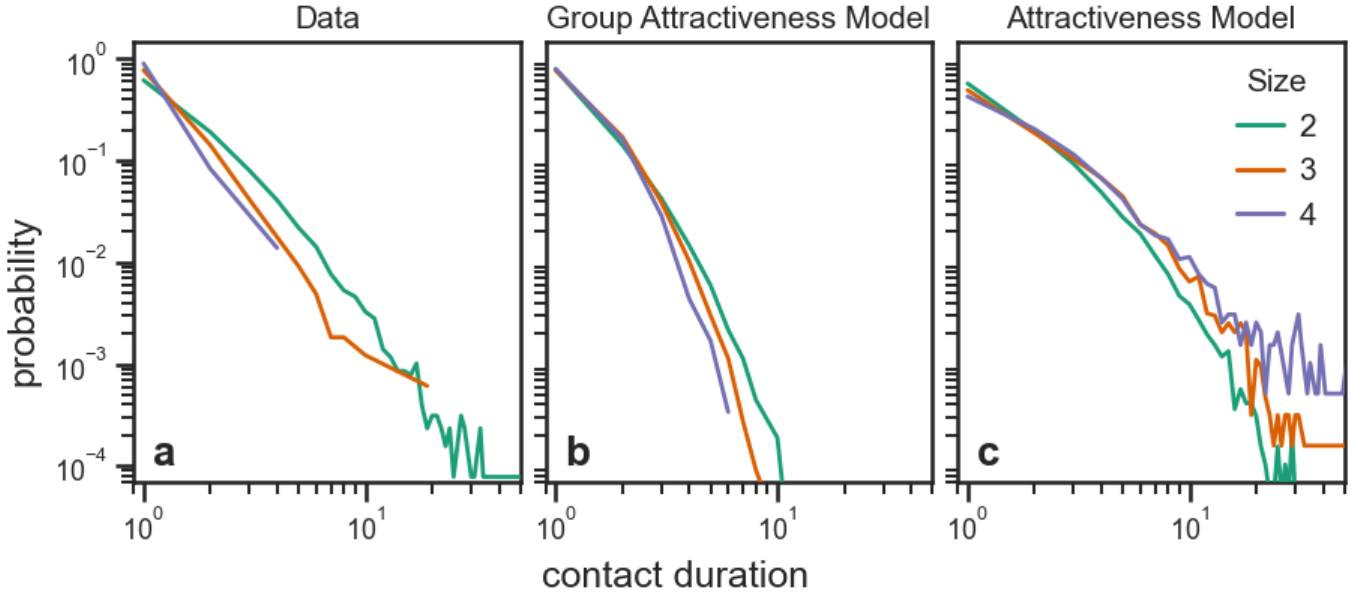}
    \caption{\textbf{Burstiness in the Hospital dataset}. We show the distributions of contact duration for groups of different sizes in the ``H'' dataset (panel \textbf{a}), as well as those predicted by the Group Attractiveness Model (panel \textbf{b}) and the Attractiveness Model (panel \textbf{c}).}
    \label{fig:rev_S4}
\end{figure}

\begin{figure}[p]
    \centering
    \includegraphics[width=\columnwidth]{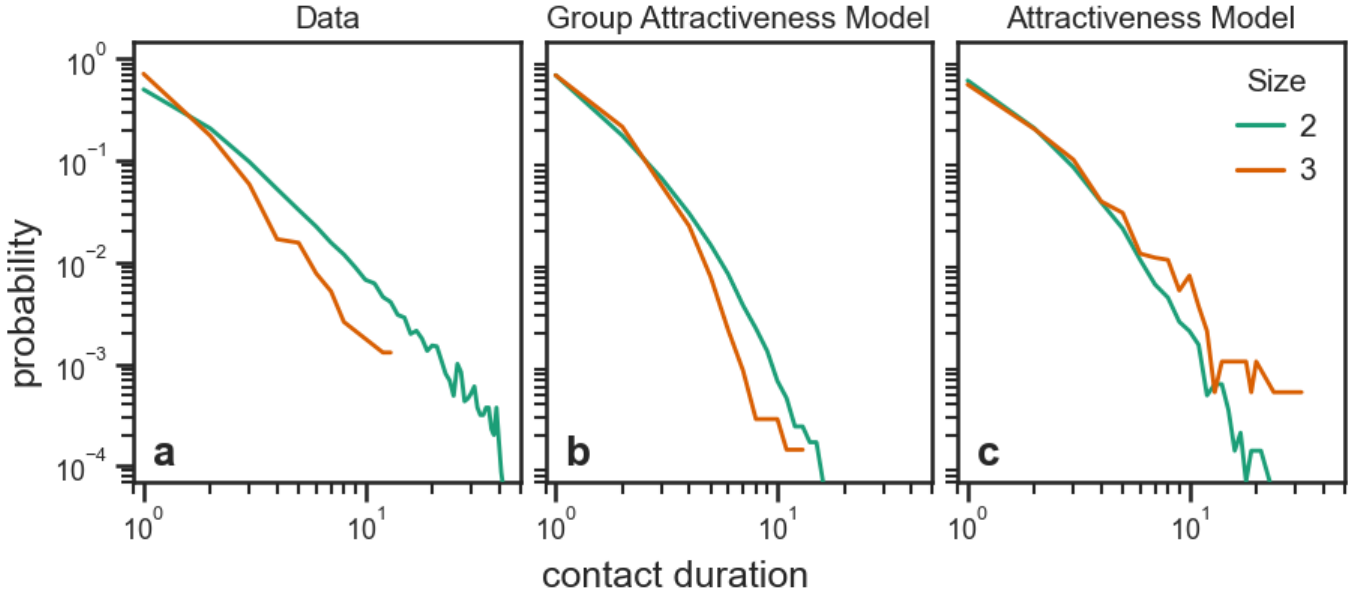}
    \caption{\textbf{Burstiness in the Malawi Village dataset}. We show the distributions of contact duration for groups of different sizes in the ``MV'' dataset (panel \textbf{a}), as well as those predicted by the Group Attractiveness Model (panel \textbf{b}) and the Attractiveness Model (panel \textbf{c}).}
    \label{fig:rev_S5}
\end{figure}

\begin{figure}[p]
    \centering
    \includegraphics[width=\columnwidth]{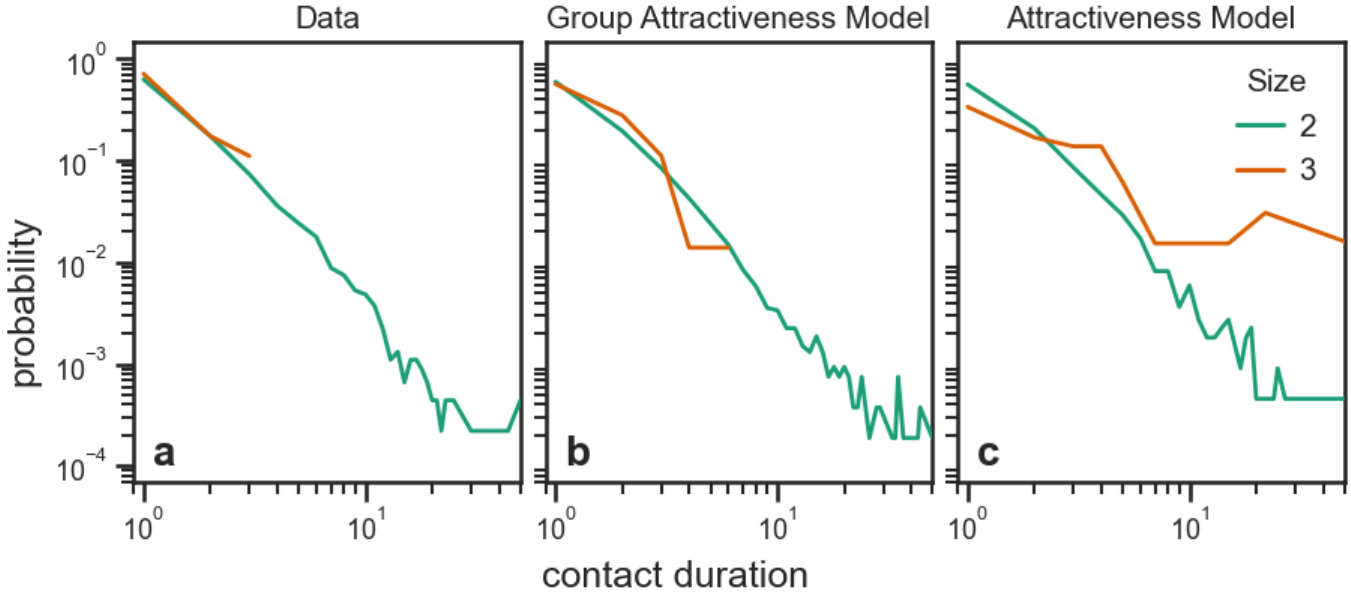}
    \caption{\textbf{Burstiness in the Workplace dataset}. We show the distributions of contact duration for groups of different sizes in the ``WP'' dataset (panel \textbf{a}), as well as those predicted by the Group Attractiveness Model (panel \textbf{b}) and the Attractiveness Model (panel \textbf{c}).}
    \label{fig:rev_S6}
\end{figure}

Panels \textbf{a} of \cref{fig:S1} to \cref{fig:rev_S6} show the distributions of contact duration for groups of different sizes in the ``HS12'', ``HS13'', ``PS'', ``C16'', ``C17'', ``SG'', ``H'', ``MV'', ``WP'', \rev{``UC'', ``MS'', and ``ES''} datasets, respectively.
In general, a hierarchy in the distributions is observed: Smaller groups have broader probability distributions compared to larger groups, meaning that smaller groups tend to last longer than larger ones.

\begin{figure}[p]
    \centering
    \includegraphics[width=\columnwidth]{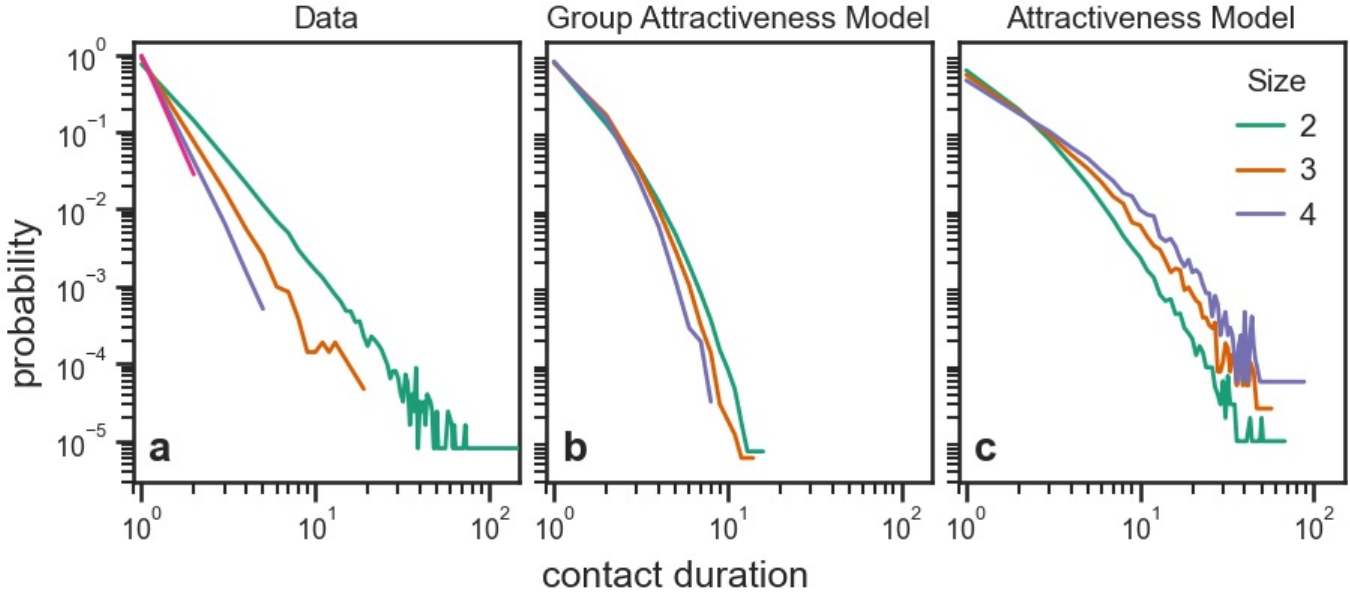}
    \caption{\rev{\textbf{Burstiness in the Elementary school dataset}. We show the distributions of contact duration for groups of different sizes in the ``ES'' dataset (panel \textbf{a}), as well as those predicted by the Group Attractiveness Model (panel \textbf{b}) and the Attractiveness Model (panel \textbf{c}).}}
    \label{fig:rev2_es_b}
\end{figure}

The emergence of broad-tailed distributions organized in a hierarchy is a feature correctly captured by the Group Attractiveness Model, as shown in Panels \textbf{b} of \cref{fig:S1} to \rev{\cref{fig:rev2_es_b}}.
Comparing data and model outcomes, we observe that the predicted distributions of contact duration are often narrower than the empirical ones.
In the ``C16'', \rev{and ``UC''} datasets, in particular, the maximum contact duration predicted is one order of magnitude smaller than the observed one.
As explained in the main text, this effect is probably due to the high density of agents in the model (for \rev{``UC''}, we have the highest density of agents, namely \rev{$N/L^2 = 3.85$}; see \cref{tab:S1}). 
In a dense environment, all agents will interact with a probability that tends to the average value of the group attractiveness (see Eq.~(2)) in the main text), weakening the impact of the agents with the highest attractiveness, which are those contributing the most to the persistence of an interaction (see also \cite{starnini2013modeling}).

Contrary to our model, the Attractiveness Model cannot reproduce the hierarchical organization of the probability distributions, as displayed in panels \textbf{c} of \cref{fig:S1} to \rev{\cref{fig:rev2_es_b}}.
In contrast with empirical observation, in the AM, larger groups of agents remain in contact longer than small groups. 
As discussed in \cite{starnini2016model}, the discrepancy with the data is related to the fact that large groups are generally more attractive than small ones, as large groups are more likely to have individuals with high attractiveness among their members.
This result marks the need to consider a higher-order mechanism of social attractiveness in order to capture burstiness at the level of group interactions.

\begin{figure}[p]
    \centering
    \includegraphics[width=\columnwidth]{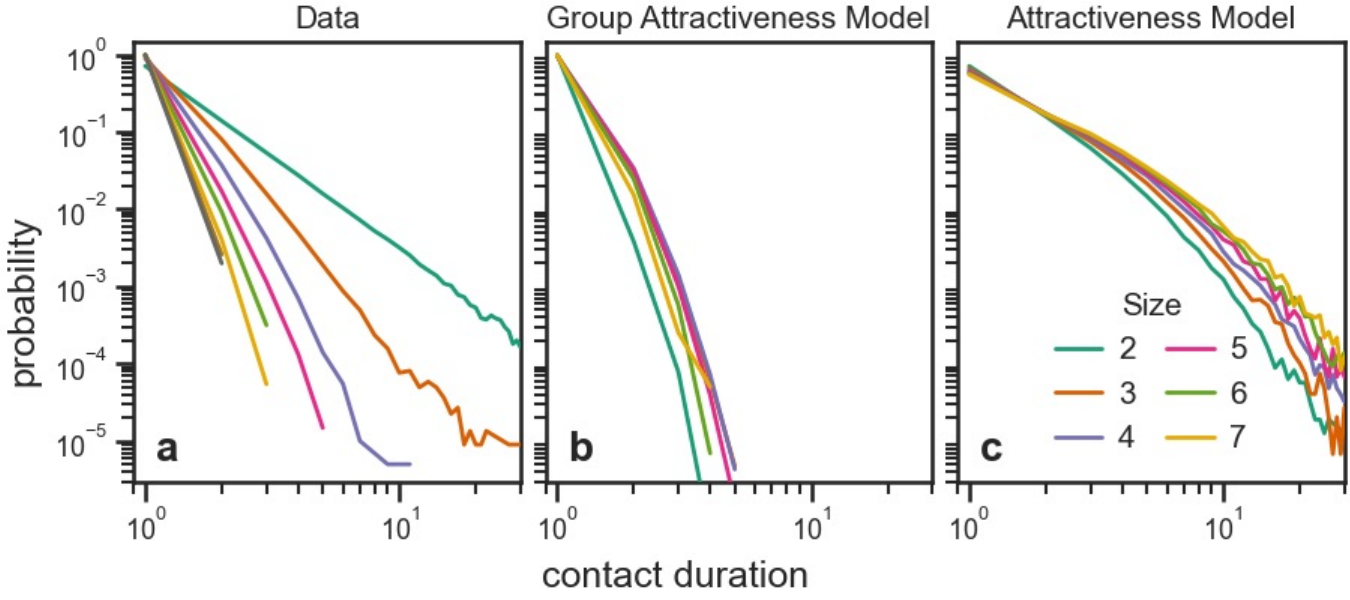}
    \caption{\rev{\textbf{Burstiness in the University dataset}. We show the distributions of contact duration for groups of different sizes in the ``UC'' dataset (panel \textbf{a}), as well as those predicted by the Group Attractiveness Model (panel \textbf{b}) and the Attractiveness Model (panel \textbf{c}).}}
    \label{fig:rev2_uc_b}
\end{figure}

\begin{figure}[p]
    \centering
    \includegraphics[width=\columnwidth]{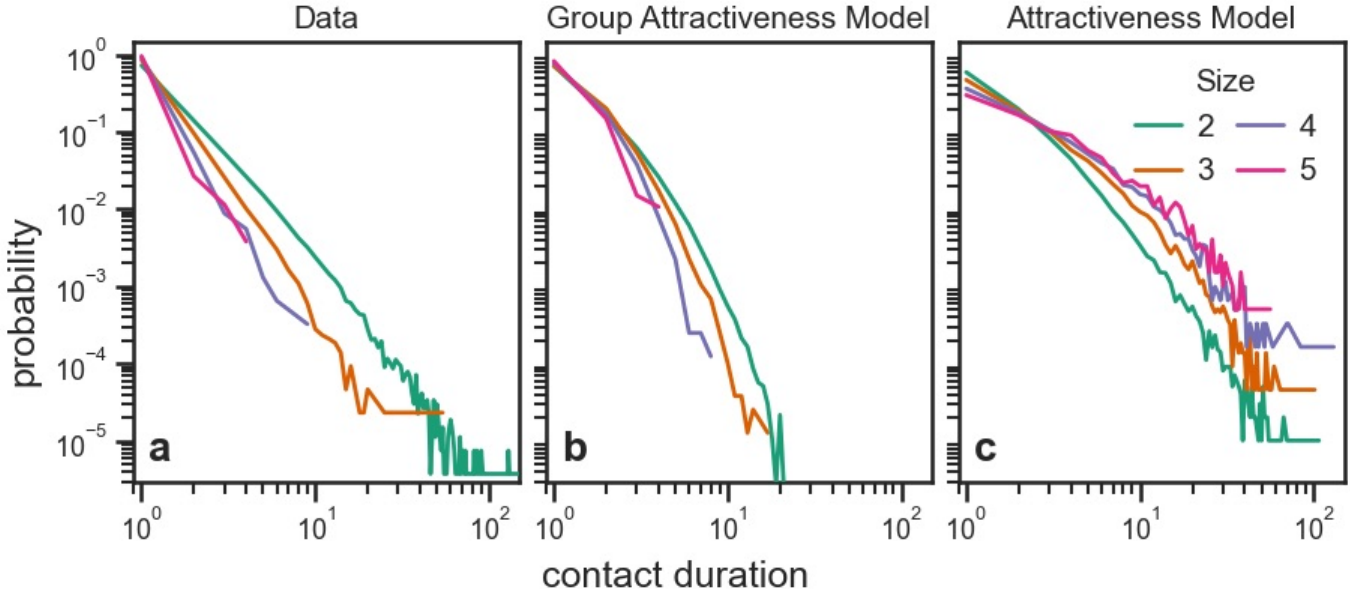}
    \caption{\rev{\textbf{Burstiness in the Middle school dataset}. We show the distributions of contact duration for groups of different sizes in the ``MS'' dataset (panel \textbf{a}), as well as those predicted by the Group Attractiveness Model (panel \textbf{b}) and the Attractiveness Model (panel \textbf{c}).}}
    \label{fig:rev2_ms_b}
\end{figure}

\clearpage

\section*{Higher-order homophily}
In many social contexts, people prefer to connect with those they perceive as similar to themselves, a characteristic known as homophily.
Previous research has measured homophily at the level of dyadic interactions, while recent studies have aimed to estimate homophilic preferences in groups of three or more individuals \cite{veldt2023combinatorial,sarker2024higher}.
In the main text, we discuss how the Group Attractiveness Model can be used to analyze higher-order homophilic patterns in face-to-face interactions, focusing on the ``HS11'' dataset.
Here, we show the results obtained for other social systems, demonstrating the nontrivial relationship between homophily and group size.

\begin{figure}[p]
    \centering
    \includegraphics[width=\textwidth]{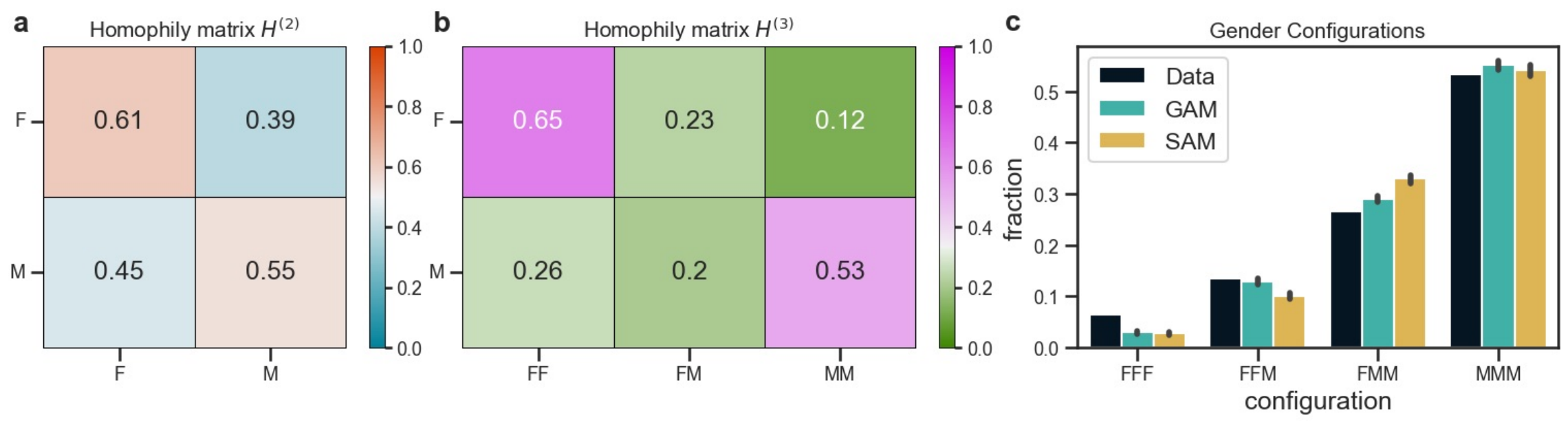}
    \caption{\textbf{Higher-order homophily in the High-school 2012 dataset}. Panels \textbf{a} and \textbf{b} display the homophily matrices $H^{(2)}$ and $H^{(3)}$, modulating the formation of groups of two and three individuals, respectively, obtained for the interactions in the ``HS12'' dataset. 
    Panel \textbf{c} shows the fraction of unique groups of size three in the different gender configurations present in the ``HS12'' dataset (black bars), together with those predicted by the Group Attractiveness Model (blue bars) and by the Social-Attractiveness Model (yellow bars).
    The value of the bars represents the average fraction over 100 simulations, while the error bars indicate the standard deviation. 
    }
    \label{fig:S6}
\end{figure}

\begin{figure}[p]
    \centering
    \includegraphics[width=\textwidth]{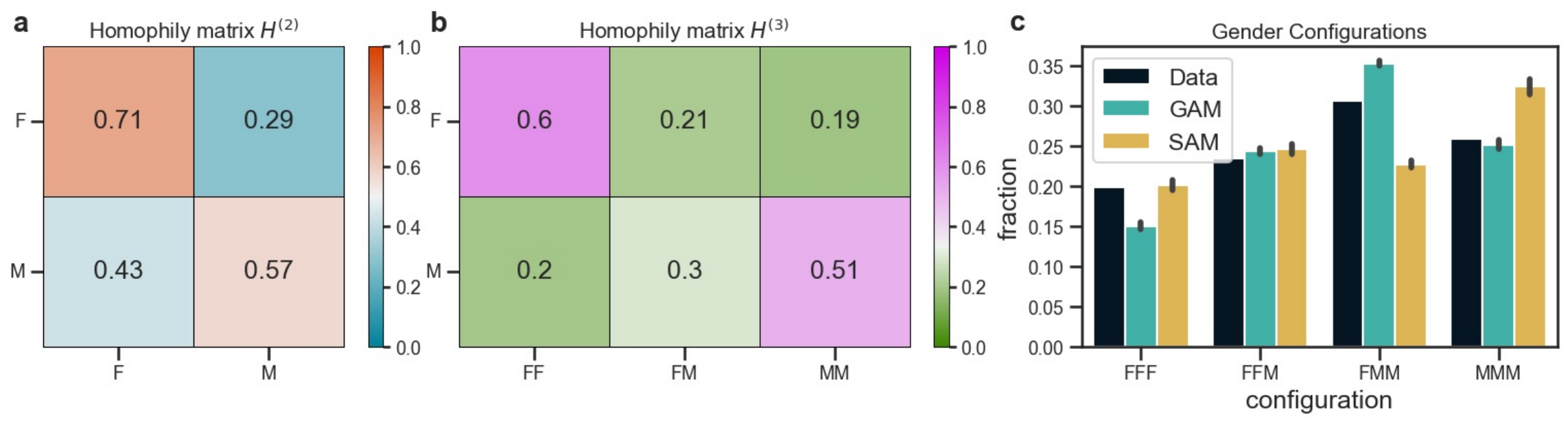}
    \caption{\textbf{Higher-order homophily in the High-school 2013 dataset}. Panels \textbf{a} and \textbf{b} display the homophily matrices $H^{(2)}$ and $H^{(3)}$, modulating the formation of groups of two and three individuals, respectively, obtained for the interactions in the ``HS13'' dataset. 
    Panel \textbf{c} shows the fraction of unique groups of size three in the different gender configurations present in the ``HS13'' dataset (black bars), together with those predicted by the Group Attractiveness Model (blue bars) and by the Social-Attractiveness Model (yellow bars).
    The value of the bars represents the average fraction over 100 simulations, while the error bars indicate the standard deviation. 
    }
    \label{fig:S7}
\end{figure}

\begin{figure}[p]
    \centering
    \includegraphics[width=\textwidth]{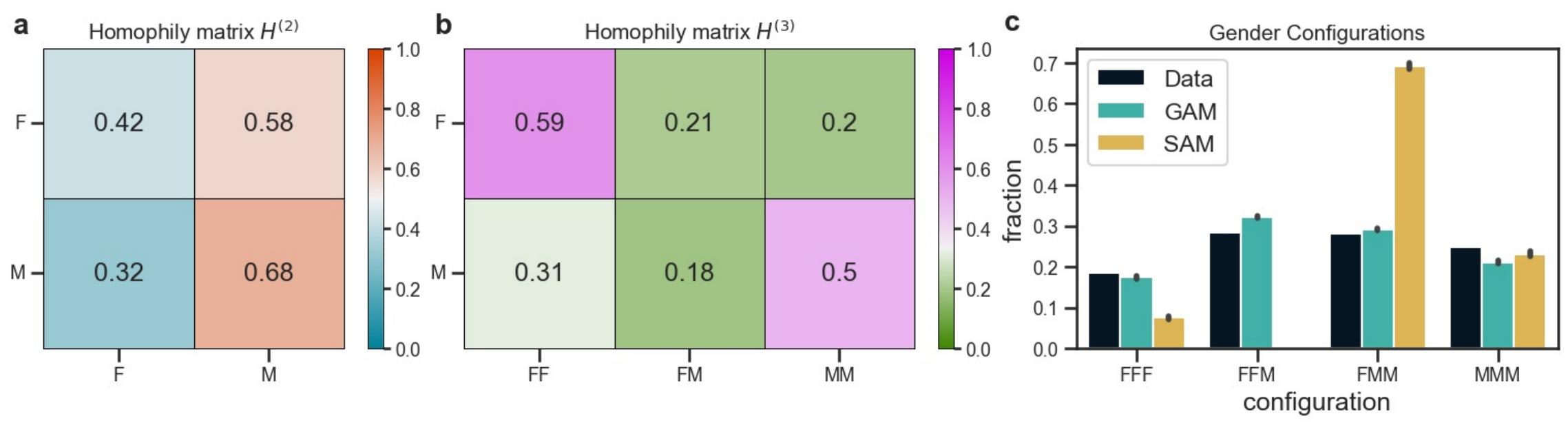}
    \caption{\textbf{Higher-order homophily in the Primary school dataset}. Panels \textbf{a} and \textbf{b} display the homophily matrices $H^{(2)}$ and $H^{(3)}$, modulating the formation of groups of two and three individuals, respectively, obtained for the interactions in the ``PS'' dataset. 
    Panel \textbf{c} shows the fraction of unique groups of size three in the different gender configurations present in the ``PS'' dataset (black bars), together with those predicted by the Group Attractiveness Model (blue bars), and by the Social-Attractiveness Model (yellow bars).
    The value of the bars represents the average fraction over 100 simulations, while the error bars indicate the standard deviation. 
    }
    \label{fig:S8}
\end{figure}

\begin{figure}[p]
    \centering
    \includegraphics[width=\textwidth]{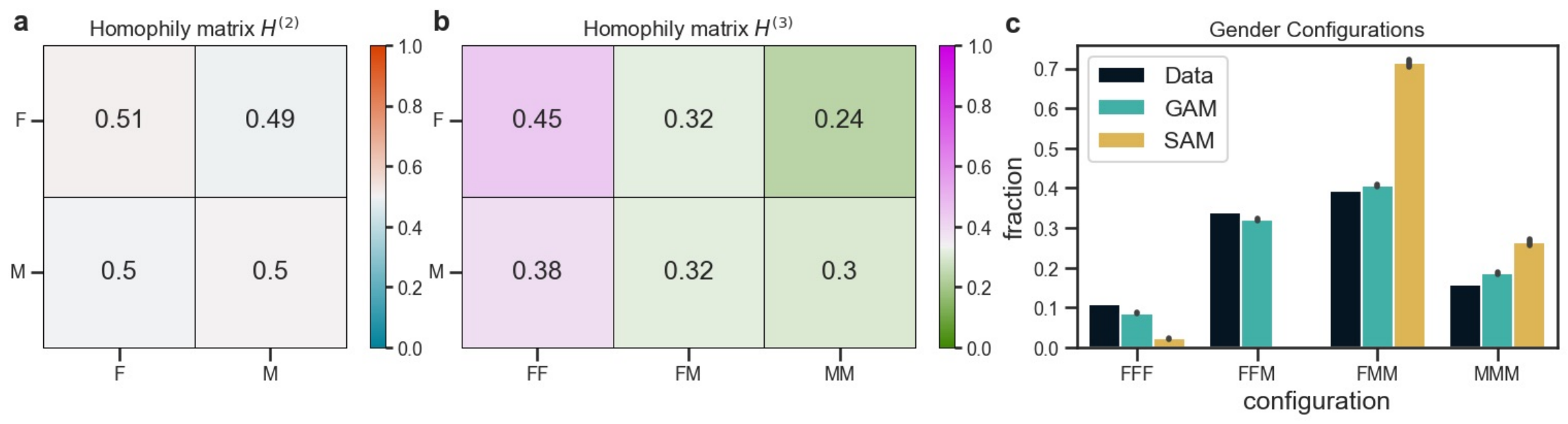}
    \caption{\textbf{Higher-order homophily in the Conference 2016 dataset}. Panels \textbf{a} and \textbf{b} display the homophily matrices $H^{(2)}$ and $H^{(3)}$, modulating the formation of groups of two and three individuals, respectively, obtained for the interactions in the ``C16'' dataset. 
    Panel \textbf{c} shows the fraction of unique groups of size three in the different gender configurations present in the ``C16'' dataset (black bars), together with those predicted by the Group Attractiveness Model (blue bars) and by the Social-Attractiveness Model (yellow bars).
    The value of the bars represents the average fraction over 100 simulations, while the error bars indicate the standard deviation. 
    }
    \label{fig:S9}
\end{figure}

\begin{figure}[p]
    \centering
    \includegraphics[width=\textwidth]{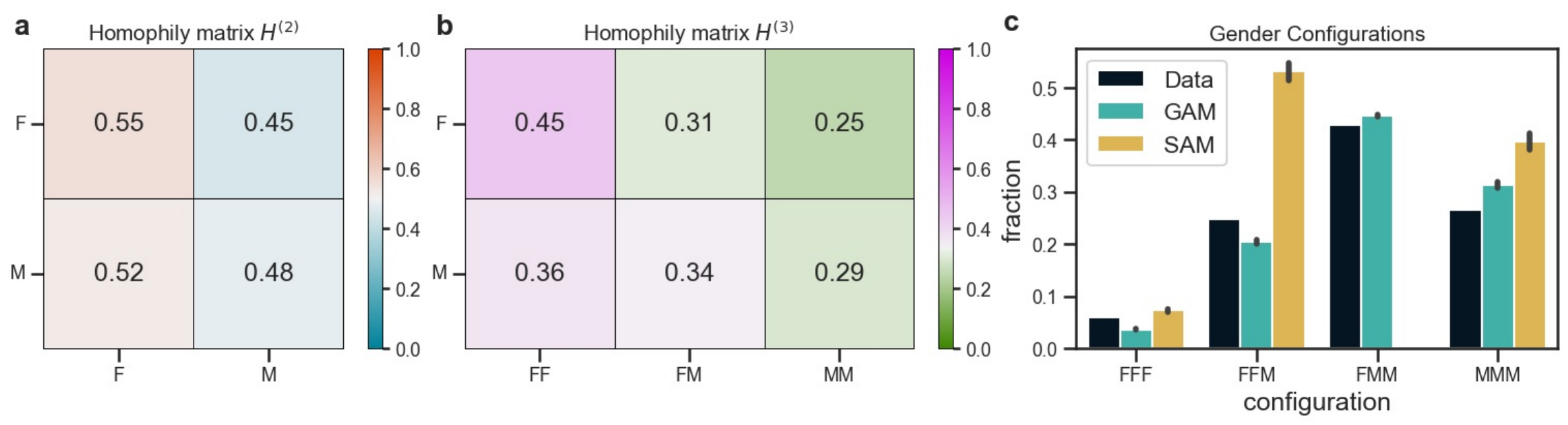}
    \caption{\textbf{Higher-order homophily in the Conference 2017 dataset}. Panels \textbf{a} and \textbf{b} display the homophily matrices $H^{(2)}$ and $H^{(3)}$, modulating the formation of groups of two and three individuals, respectively, obtained for the interactions in the ``C17'' dataset. 
    Panel \textbf{c} shows the fraction of unique groups of size three in the different gender configurations present in the ``C17'' dataset (black bars), together with those predicted by the Group Attractiveness Model (blue bars) and by the Social-Attractiveness Model (yellow bars).
    The value of the bars represents the average fraction over 100 simulations, while the error bars indicate the standard deviation. 
    }
    \label{fig:S10}
\end{figure}

Panels \textbf{a} and \textbf{b} of \cref{fig:S6} to \cref{fig:S10} display the homophily matrices $H^{(2)}$ and $H^{(3)}$, modulating the interactions in pairs and triples, obtained for the interactions in the ``HS12'', ``HS13'', ``PS'', ``C16'', ``C17'' datasets, respectively (details on how to evaluate these matrices are provided in the Methods of the main text).
We observe a variety of scenarios, complementing the one described in the main text for ``HS11''.
Specifically, in ``HS12'' and ``HS13'', both women and men have homophilic tendencies: Each individual prefers to interact with those of the same gender, as signaled by the difference in the interaction probabilities.
Here, homophilic preferences do not change based on how many individuals interact together: Both in pairs and triples, women and men display homophilic tendencies, i.e., $h_{00} > h_{01}$ and $h_{11} > h_{10}$, in dyadic interactions, while $h_{000} > h_{001} + h_{011}$ and $h_{111} > h_{100} + h_{101}$ in triadic ones.
As shown in the main text for ``HS11'', consistency in homophilic preferences across group sizes is not a general rule.
In ``PS'', for instance, we observe that men tend to be homophilic at both levels of pairs and triples, whereas women have discordant behavior: They prefer to interact with men in dyadic interactions ($h_{00} < h_{01}$), and more exclusively with women in triadic ones ($h_{000} > h_{001} + h_{011}$).
Finally, ``C16'' and ``C17'' exhibit a more neutral scenario where face-to-face interactions are not strongly determined by homophily. 
At the level of groups of size two, the interaction probabilities are almost the same, signaling the absence of a relevant homophilic tendency.
At the level of groups of size three, women display a weak preference for interacting exclusively with other women ($h_{000}$ is the largest probability), while men are neutral.
Overall, these results showcase the multifaceted nature of higher-order homophily.

Panels \textbf{c} shows the fraction of unique groups of size three in the different gender configurations present in the various datasets (black bars), together with those predicted by the GAM (blue bars) and the Social-Attractiveness Model (SAM) proposed in \cite{oliveira2022group} (yellow bars).
The overall better performance of the GAM, based on higher-order homophily, compared to the SAM, which deals with homophily only at the level of dyadic interactions, proves the importance of considering homophily in groups of various sizes.  
In ``PS'', ``C16'', ``C17'', and ``HS11'' in the main text, the SAM largely overestimates one of the more heterophilic groups, i.e., FFM and FMM, while predicting the other to be zero.
Conversely, our model provides significantly better predictions of the empirical mixing patterns in these social systems.  
Even in those datasets where the SAM provides more reliable predictions, i.e., in ``HS12'' and ``HS13'', GAM shows either better or comparable performances, corroborating the benefit of considering homophily at the level of groups.

\clearpage

{\color{crimson}
\section*{Group overlap in face-to-face contacts}
In both data and models, groups active at a given time can overlap, meaning that they can share some of their members.
Importantly, one group cannot be a subgroup of another, as this is forbidden by the GAM and cannot occur in the data and the pairwise models, as we infer groups as the maximal cliques appearing at a given time (see Methods)
Here, we discuss the extent to which groups in the data and the GAM overlap, i.e., they share some of their members.
We quantify the degree of group overlap as follows:
For each time step $t$ where at least two groups are active, we built a graph where nodes represent the active groups, while links represent whether two groups overlap or not, i.e., we put a link if the groups share any of their members.
We evaluate the degree of overlap at time $t$ as the density of this graph, namely as the number of links in the graph, namely the number of overlaps, divided by the maximal number of links, that is the maximal possible number of overlaps.
Finally, we take the average over all time steps as the measure of group overlap of the data/model. 

\begin{figure}[h]
    \centering
    \includegraphics[width=.5\columnwidth]{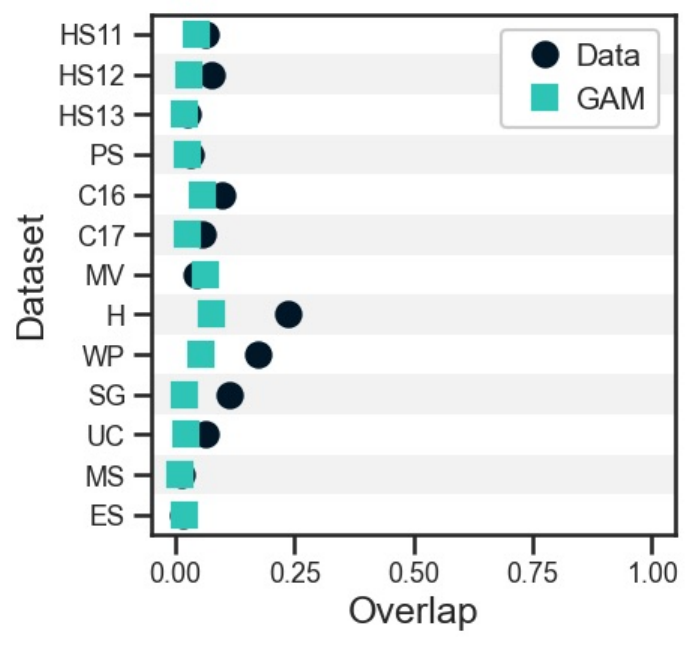}
    \caption{\rev{\textbf{Group overlap in face-to-face interactions}. 
    The degree overlap varies across social systems (black circles). 
    Despite not explicitly preventing it, the Group Attractiveness Model shows low levels of group overlap (blue squares).}}
    \label{fig:overlap_SM}
\end{figure}

We measure group overlap in all datasets and their respective simulations.
Each dataset has a certain degree of overlap, though this is generally low, less than $10\%$ in most datasets (see \cref{fig:overlap_SM}).
This is consistent across different social systems and sensing technologies.
Remarkably, the overlap is larger in the Hospital, Workplace, and Science Gallery datasets, suggesting that context may affect the way group organize in space.
Despite the GAM does not forbid group overlap explicitly, we observe a low degree of overlap for all datasets considered.
We also see that the level of group overlap in the GAM is comparable to that observed in the data, with the exception of the Hospital, Workplace, and Science Gallery datasets, where the GAM does not show the large degree of overlap found in the data.
Overall, this analysis suggests that our findings are likely to be unaffected by group overlap, as in most datasets the level of overlap is not substantial and is comparable to the value obtained with the GAM.
}

\clearpage

\def\bibsection{\section*{Supplementary References}} 
\bibliography{supp-biblio}